\documentstyle[prd,aps,eqsecnum,preprint,tighten,floats]{revtex}

\def\senk#1{\bbox{#1}_\perp}
\def\g{\gamma}
\def\ha{{1\over 2}}
\def\L{\Lambda}
\def\ub#1{\underline{#1}}
\def\ths{\thinspace}
\def\psibar{\overline{\psi}}
\def\etabar{\overline{\eta}}
\def\del{\partial}

\def\eg{{\it e.g.}}
\def\pri{\prime}

\def\a{\alpha}
\def\b{\beta}
\def\d{\delta}
\def\e{\epsilon}
\def\kd3{\delta^{(3)}}

\def\ran{\rangle}

\def\ket#1{|#1\ran}

\begin{document}

\draft

\preprint{\vbox{\hfill SLAC-PUB-8920 \\
          \vbox{\hfill UMN-D-01-6}  \\
          \vbox{\hfill SMUHEP/01-06 }
          \vbox{\vskip0.3in}
          }}

\title{Exact solutions to Pauli--Villars-regulated field theories%
\footnote{Work supported in part by the Department of Energy
under contract numbers DE-AC03-76SF00515, DE-FG02-98ER41087,
and DE-FG03-95ER40908.}}

\author{Stanley J. Brodsky}
\address{Stanford Linear Accelerator, Stanford University, 
Stanford, California 94309}

\author{John R. Hiller}
\address{Department of Physics, University of Minnesota-Duluth, 
Duluth, Minnesota 55812}

\author{Gary McCartor}
\address{Department of Physics, Southern Methodist University, 
Dallas, TX 75275}

\date{\today}

\maketitle

\begin{abstract} 
We present a new class of quantum field theories
which are exactly solvable.  The theories are generated
by introducing Pauli--Villars fermionic and bosonic fields with
masses degenerate with the physical positive metric fields.
An algorithm is given to compute the spectrum and corresponding
eigensolutions.  We also
give the operator solution for a particular case and use it to illustrate
some of the tenets of light-cone quantization.  Since the solutions
of the solvable theory contain ghost quanta, these theories are
unphysical.  However, we also discuss how perturbation theory in the
difference between the masses of the physical and Pauli--Villars particles 
could be developed, thus generating physical theories.  The existence of
explicit solutions of the solvable theory also allows one to study
the relationship between the equal-time and light-cone vacua
and eigensolutions.
\end{abstract}

\narrowtext

\section{Introduction}

Exact solutions to quantum field theories in physical space-time with
non-trivial interactions are rare.  In this paper we shall show how one
can obtain the complete eigenspectrum and eigensolutions of a quantum
field theory of interacting massive fermions and bosons in 3+1 space-time
dimensions.  The basic format of the solvable theory is the conventional
Yukawa theory with
$g\phi\psibar\psi$ interactions, accompanied by negative-metric
Pauli--Villars (PV) boson and fermion fields with masses degenerate with the
physical quanta.  An algorithm is then given which generates the complete
eigenspectrum and the corresponding eigensolutions, with respect to
a light-cone-quantized Fock basis.

Since the solutions of the
solvable theory contain ghost quanta, these theories are unphysical.
However, we will discuss how perturbation theory in the difference
between the masses of the physical and PV particles can be developed, thus
ultimately generating physical theories in which the wave functions allow 
one to compute space-like and time-like form factors and other quantities 
of phenomenological interest.  Conversely, the exact solutions
provide boundary conditions for the wave functions of the physical theory
in the limit of degenerate masses.  The explicit solutions of the
solvable theory also allow one to study the relationship
between the equal-time and light-cone vacua and eigensolutions, and they
display properties due to covariance, such as
light-cone spin conservation, which are characteristic of physical
nonperturbative eigensolutions.  In addition, such solutions provide
important checks of computer codes for discretized light-cone quantization
(DLCQ)~\cite{PauliBrodsky,DLCQreview} of nondegenerate theories~\cite{bhm3}.

Pauli--Villars regularization~\cite{pv} is an important method
for regulating the ultraviolet divergences of light-cone Hamiltonian
theories.  We have previously shown that the use of 
PV regulation provides a correct renormalization of DLCQ,
for Yukawa theory at least to one loop; in contrast, a momentum
cutoff of DLCQ does not preserve the chiral properties of the
theory~\cite{bhm1}.  We have also made a number of studies
showing the practicality of using PV regulation in 3+1 nonperturbative
DLCQ calculations~\cite{bhm1,bhm2,bhm3}.  In an important development, Paston
and Franke~\cite{pf}, and Paston, Franke and Prokhvatilov~\cite{pfp} have
now shown that regulation with the correct combination
of PV fields always gives perturbative agreement with Feynman theory, and
they have given a complex set of rules for deciding which set of PV fields
are sufficient to regulate a given theory at all orders.

In the present paper we will show that if a theory is regulated with PV
fields and the masses of the PV fields are set equal to the masses of the
physical fields, the resulting theory is easy to solve.
After some discussion of PV-regulated Yukawa theory in Sec.~\ref{sec:PVYukawa},
we give the general procedure for finding eigenvalues and
eigenvectors in Sec.~\ref{sec:Exact}\@.  We then give in
Sec.~\ref{sec:OpSoln} an operator solution for Yukawa theory and use it
to find the relation between the light-cone basis states and the
equal-time basis states.  We show that, not only is the light-cone
perturbative vacuum equal to the physical vacuum while the equal-time
perturbative vacuum is not, but that all the eigenstates are much simpler
when expressed in the light-cone representation than when expressed in
the equal-time representation.  In Sec.~\ref{sec:SeveralPV} we show
that the procedure works for the case of several PV fields of the same
type and give an explicit example.

Since the masses of the ghost metric quanta are degenerate with the masses
of physical particles, the solutions of the solvable theories will violate
unitarity,  and they are thus not physical.  Since the exact
solutions exist for any value of the coupling constant, one can
construct the solution of a theory with large values of the PV masses as
a perturbative expansion, not in powers of the coupling constant, but in
powers of the difference between the PV masses and the physical masses.
Such a perturbation theory would require the expansion parameter to have
large values, so its practical utility will depend on the analytic
properties of the solution in the mass differences.  The calculation of
scattering matrix elements and other physical quantities as a
perturbation theory in powers of the mass differences involves
polynomials in the coupling constant of no higher order than the order of 
the expansion in the masses.  Therefore, although it might at first seem
otherwise, low orders of the new series will not contain information from
higher order Feynman graphs.  The solutions may be of use in studying
various general properties of quantum field theories even if the
convergence of the new perturbation series is not very good.  We do know
that the solutions have at least one use: we have used them to help debug
the computer code used in the calculations of~\cite{bhm3}.  Additional
conclusions and applications are discussed in Sec.~\ref{sec:Conclusions}\@.  
Our light-cone conventions and definitions are collected in an appendix.

\section{Pauli--Villars Regularization of Yukawa Theory}
\label{sec:PVYukawa}

We begin our discussion with the light-cone quantization of Yukawa theory
in $3+1$ dimensions.  In order to keep the notation
as simple as possible, we shall first introduce
just one PV boson and one PV fermion although this is
not sufficient to regulate the full Yukawa theory; the results of
Paston and Franke~\cite{pf} show that one bosonic PV field and two PV
fermion fields will regulate the theory in such a way that it is
perturbatively equivalent to Feynman theory.  However, for the purposes of
this section, one can omit fermion loops
(the theory studied in~\cite{bhm3}), or introduce an additional transverse
momentum cutoff.  Paston {\em et al}.~\cite{pfp2} have suggested that one
bosonic PV field and one PV fermion field plus
a transverse cutoff regulates the theory in such a way as to generate only
mass, coupling constant, and wave function renormalization.

Taking the physical fields to be $\psi_1$ and $\phi_1$ and the PV
(negative-metric) fields to be $\psi_2$ and $\phi_2$, the action becomes
\begin{eqnarray} 
S=\int d^4x&&
\Biggl[\ha(\del_\mu\phi_1)^2-\ha\mu_1^2\phi_1^2
-\ha(\del_\mu\phi_2)^2+\ha\mu_2^2\phi_2^2 
\nonumber \\
+&& {i\over2}\Bigl(\psibar_1\g^\mu\del_\mu-(\del_\mu\psibar_1)\g^\mu\Bigr)
     \psi_1-m_1\psibar_1\psi_1
-{i\over2}\Bigl(\psibar_2\g^\mu\del_\mu-(\del_\mu\psibar_2)\g^\mu\Bigr)
      \psi_2+m_2\psibar_2\psi_2 
\nonumber \\  
-&& g\phi\psibar\psi\Biggr]\,,					
\end{eqnarray} 
where the Yukawa three-point interaction is expressed in terms of
\begin{equation}
\psi \equiv {1 \over \sqrt{2}}(\psi_1 + \psi_2)\,,  \quad
                \phi \equiv {1 \over \sqrt{2}}(\phi_1 + \phi_2)\,.
\end{equation}
For simplicity, we have not considered here a $\phi^4$ term; it is,
however, easily included.  We also define
\begin{equation}
\eta \equiv {1 \over \sqrt{2}}(\psi_1 - \psi_2)\,,\quad
\zeta \equiv {1 \over \sqrt{2}}(\phi_1 - \phi_2)\,.
\end{equation} 
Notice that $\eta$ and $\zeta$ are zero-norm fields.  In
terms of these fields we have
\begin{eqnarray} 
S=\int d^4x&& \Biggl[\ha(\del_\mu\phi\del^\mu\zeta
+\del_\mu\zeta\del^\mu\phi)-\ha(\ha(\mu_1^2 +\mu_2^2))(\phi\zeta +
\zeta\phi) +\ha(\ha(\mu_2^2 - \mu_1^2))(\phi^2 + \zeta^2) 
\nonumber \\
 +&&{i\over2}\Bigl(\psibar\g^\mu\del_\mu-(\del_\mu\psibar)\g^\mu\Bigr)\eta
  +{i\over2}\Bigl(\etabar\g^\mu\del_\mu-(\del_\mu\etabar)\g^\mu\Bigr)\psi
\nonumber \\
 -&&\ha(m_1 + m_2)(\psibar\eta +\etabar\psi) + 
      \ha(m_2 - m_1)(\psibar\psi + \etabar\eta) 
\nonumber \\  
 -&&g\phi\psibar\psi\Biggr]\,.					
\end{eqnarray}

The light-cone Hamiltonian $P^-$ can be constructed using the methods
of~\cite{mr}:
\begin{equation} 
P^-=\ha\int dx^- d\senk{x} \ths\ths T^{+-}\,,					
\end{equation}
\begin{eqnarray} 
T^{+-}&&= -\del_i\phi\del^i\zeta -
\del_i\zeta\del^i\phi+ (\mu^2 + \ha\delta^2)(\phi\zeta + \zeta\phi) -
\ha\delta^2(\phi^2 + \zeta^2) 
\nonumber \\ 
&& -2i(\psi_-^\dagger\del_-\eta_- - (\del_-\psi_-)^\dagger\eta_- +
\eta_-^\dagger\del_-\psi_- - (\del_-\eta_-)^\dagger\psi_-) 
\nonumber \\
&&-i(\psi_-^\dagger \a^i \del_i\eta_+ + \psi_+^\dagger \a^i \del_i\eta_-
- (\del_i\psi_+)^\dagger\a^i\eta_- - (\del_i\psi_-)^\dagger\a^i\eta_+
\nonumber \\ 
&&+ \eta_-^\dagger \a^i \del_i\psi_+ + \eta_+^\dagger \a^i
\del_i\psi_- - (\del_i\eta_+)^\dagger\a^i\psi_- -
(\del_i\eta_-)^\dagger\a^i\psi_+) 
\nonumber \\ 
&&+ 2(m + \ha\Delta)(\psi_+^\dagger\g^0\eta_- + \psi_-^\dagger\g^0\eta_+ +
\eta_+^\dagger\g^0\psi_- + \eta_-^\dagger\g^0\psi_+) 
\nonumber \\ 
&&- \Delta (\psi_+^\dagger\g^0\psi_- + \psi_-^\dagger\g^0\psi_+ +
\eta_+^\dagger\g^0\eta_- + \eta_-^\dagger\g^0\eta_+) 
\nonumber \\  
&& +2 g \phi (\psi_+^\dagger\g^0\psi_- + \psi_-^\dagger\g^0\psi_+)\,,
\end{eqnarray} 
where we have defined
\begin{equation}
   \mu^2\equiv \mu_1^2\,, \quad
   \delta^2 \equiv \mu_2^2 - \mu_1^2\,, \quad
   m \equiv m_1\,, \quad
   \Delta \equiv m_2 - m_1\,,
\end{equation} 
and $\a^i\equiv\g^0\g^i$ are the original Dirac matrices.
The fields $\psi_-$ and $\eta_-$ are nondynamical and must be eliminated
via the constraint relations; these take the form
\begin{equation}
   2i\del_-\eta_-=\Bigl[ -i\a^i\del_{i} + (m +
\ha\Delta)\g^0\Bigr]\eta_+  + \Bigl[-\ha\Delta\g^0
+g\g^0\phi\Bigr]\psi_+  	
\label{eometa}				
\end{equation} 
and
\begin{equation}
   2i\del_-\psi_-=\Bigl[ -i\a^i\del_{i} + (m +
\ha\Delta)\g^0\Bigr]\psi_+  + \Bigl[-\ha\Delta\g^0\Bigr]\eta_+\,.
\label{eompsi}				
\end{equation}
The mode expansions are
\begin{eqnarray}
\psi_+(\ub{x})&=&{1\over\sqrt{16\pi^3}} \sum_{s} \int d\ub{k} \; \chi_s \;
\Bigl[ b_{s,\ub{k}} e^{-i\ub{k}\cdot\ub{x}}+
d^\dagger_{-s,\ub{k}}e^{+i\ub{k}\cdot\ub{x}}\Bigr]\,, \\				
\eta_+(\ub{x})&=&{1\over\sqrt{16\pi^3}} \sum_{s} \int d\ub{k} \; \chi_s \;
\Bigl[
\beta_{s,\ub{k}} e^{-i\ub{k}\cdot\ub{x}}+
\delta^\dagger_{-s,\ub{k}}e^{+i\ub{k}\cdot\ub{x}}\Bigr]\,,  \\
\phi(\ub{x})&=&{1\over\sqrt{16\pi^3}}  \int d\ub{q} {1\over\sqrt{q^+}}\;
\Bigl[ a_{\ub{q}} e^{-i\ub{q}\cdot\ub{x}}+ a^\dagger_{\ub{q}}
e^{+i\ub{q}\cdot\ub{x}}\Bigr]\,,	\\				
\zeta(\ub{x})&=&{1\over\sqrt{16\pi^3}}  \int d\ub{q} {1\over\sqrt{q^+}} \;
\Bigl[
\alpha_{\ub{q}} e^{-i\ub{q}\cdot\ub{x}}+
\alpha^\dagger_{\ub{q}} e^{+i\ub{q}\cdot\ub{x}}\Bigr]\,.
\label{ichi}					
\end{eqnarray}
The integration measure is $d\ub{k}= d^2k_\perp dk^+$.  The
light-cone four-spinors $\chi_s$ are defined in the Appendix.
The canonical commutation relations derived from the Lagrangian are
\begin{eqnarray}
\Bigl\{\psi_{1_{+\a}}(x^+,\ub{x}),
    \psi^\dagger_{1_{+\b}}(x^+,\ub{x}^\prime)\Bigr\}
&=&(\L_+)_{\a\b}\kd3(\ub{x}-\ub{x}^\prime)\,,  \\					
\Bigl\{\psi_{2_{+\a}}(x^+,\ub{x}),
    \psi^\dagger_{2_{+\b}}(x^+,\ub{x}^\prime)\Bigr\}
&=&-(\L_+)_{\a\b}\kd3(\ub{x}-\ub{x}^\prime)\,,	\\				
\Bigl[\phi_1(x^+,\ub{x}),\del^+\phi_1(x^+,\ub{x}^\prime)\Bigr]
&=&i\kd3(\ub{x}-\ub{x}^\prime)\,,	\\				
\Bigl[\phi_2(x^+,\ub{x}),\del^+\phi_2(x^+,\ub{x}^\prime)\Bigr]
&=&-i\kd3(\ub{x}-\ub{x}^\prime)\,,				
\end{eqnarray} 
the others being zero.  These are realized by the Fock
space relations
\begin{equation}
\{b_{s,\ub{k}},\beta^\dagger_{s^\pri,\ub{k}^\pri}\}=
\{d_{s,\ub{k}},\delta^\dagger_{s^\pri,\ub{k}^\pri}\}=
\d_{ss^\prime}\kd3(\ub{k}-\ub{k}^\pri) \ths,					
\end{equation}
\begin{equation}
[a_{\ub{q}},\alpha^\dagger_{\ub{q}^\pri}]=\kd3(\ub{q}-\ub{q}^\pri)\ths.		
\end{equation} 
All others are zero, including
\begin{equation}
\{b,b^\dagger\}=\{d,d^\dagger\}=\{\beta,\beta^\dagger\}
     =\{\delta,\delta^\dagger\}
     =[a,a^\dagger]=[\alpha,\alpha^\dagger]=0.					
\end{equation}

We can now (following~\cite{mr}) give $P^-$.  We write
\begin{equation} 
P^-\equiv P^-_{(0)}+gP^-_{(1)}\,,					
\end{equation} 
where
\begin{eqnarray} 
P^-_{(0)}=&&\int d\ub{q}
\left[ {\senk{q}^{\ths2}+(\mu^2 + \delta^2/2)\over q^+}\right]
(a^\dagger_{\ub{q}}\a_{\ub{q}} + \a^\dagger_{\ub{q}}a_{\ub{q}})
\nonumber \\ 
&&+\sum_s\int d{\ub{k}}
\left[ {\senk{k}^{\ths2}+(m^2 + D^2/2)\over k^+}\right]
\Bigl(b^\dagger_{s,\ub{k}}\b_{s,\ub{k}}+\b^\dagger_{s,\ub{k}}b_{s,\ub{k}}
  +d^\dagger_{s,\ub{k}}\d_{s,\ub{k}} +
\d^\dagger_{s,\ub{k}}d_{s,\ub{k}}\Bigr)
\nonumber \\ 
&&+\int d{\ub{q}}
\left[ {-\delta^2\over 2q^+}\right] (a^\dagger_{\ub{q}}a_{\ub{q}} +
\a^\dagger_{\ub{q}}\a_{\ub{q}})
\\ 
&&+\sum_s\int d\ub{k}
\left[ {-D^2\over 2k^+}\right]
\left( b^\dagger_{s,\ub{k}}b_{s,\ub{k}}+\b^\dagger_{s,\ub{k}}\b_{s,\ub{k}}
  +d^\dagger_{s,\ub{k}}d_{s,\ub{k}} +
\d^\dagger_{s,\ub{k}}\d_{s,\ub{k}}\right)\,, \nonumber
\end{eqnarray}
and $D^2 \equiv m_2^2 - m_1^2$.  The
first-order interaction Hamiltonian separates naturally into two pieces,
describing boson emission with and without a spin flip:
\begin{equation} 
P^-_{(1)}\equiv V_{\rm flip}+V_{\rm noflip}\,.
\end{equation}
We have found it convenient to express the spin-flip part
of $P^-_{(1)}$ in terms of transverse ``polarization'' vectors
\begin{equation}
\bbox{\e}_{\perp,+1}\equiv{-1\over\sqrt{2}}(1,i)\,,  \qquad\qquad
  \bbox{\e}_{\perp,-1}\equiv {1\over\sqrt{2}}(1,-i)\,,
\end{equation} 
satisfying
\begin{equation} 
{\e^i_\l}^*\e^i_{\l^\prime}=\d_{\l\l^\prime}\,,
\qquad\quad
\sum_\l \e^i_\l{\e^j_\l}^*=\d^{ij}\,, \quad\qquad
\e^i_\l\e^i_{\l^\prime}={\e^i_\l}^*{\e^i_{\l^\prime}}^*
                                     =-\d_{\l,-\l^\prime}\,.			
\end{equation}
Of course, the boson has no such degree of freedom; this
is merely a way of writing the Hamiltonian more compactly.  Other useful
relations satisfied by the $\bbox{\e}_{\perp,\l}$ include:
\begin{eqnarray}
\bbox{\e}_{\perp,\l}^{\ths*}&= &- \bbox{\e}_{\perp,-\l}\,,	\\				
v^i\chi^\dagger_s \b\a^i\chi_{s^\pri} &=&\sqrt{2}\ths
\bbox{\e}_{\perp,2s}^{\ths*}\cdot \senk{v} \ths\d_{s,-s^\pri}\,,					
\end{eqnarray}
where $\senk{v}$ is any transverse vector.
We have
\begin{eqnarray} 
V_{\rm flip}={1\over \sqrt{8\pi^3}} &&\sum_{s} \int d\ub{k} d\ub{l}d\ub{q}
\ths\ths{ ({\bbox{\e}_{\perp,2s}}^{\ths *}\cdot\senk{l}) \over l^+\sqrt{q^+} }
\nonumber \\
\times\Bigl\{ &&b^\dagger_{s,\ub{k}}b_{-s,\ub{l}}a_{\ub{q}}
	\ths\kd3(\ub{k}-\ub{l}-\ub{q})
+b^\dagger_{s,\ub{k}}d^\dagger_{s,\ub{l}}a_{\ub{q}}
	\ths\kd3(\ub{q}-\ub{k}-\ub{l}) \\
+&&b^\dagger_{s,\ub{k}}b_{-s,\ub{l}}a^\dagger_{\ub{q}}
	\ths\kd3(\ub{k}+\ub{q}-\ub{l})
-d^\dagger_{s,\ub{l}}d_{-s,\ub{k}}a_{\ub{q}}
	\ths\kd3(\ub{k}+\ub{q}-\ub{l})
\nonumber \\ 
+&&d_{-s,\ub{k}}b_{-s,\ub{l}}a^\dagger_{\ub{q}}
	\ths\kd3(\ub{k}+\ub{l}-\ub{q})
-d^\dagger_{s,\ub{l}}d_{-s,\ub{k}}a^\dagger_{\ub{q}}
	\ths\kd3(\ub{k}-\ub{l}-\ub{q}) \Bigr\}+{\rm h.c.}\,,
\nonumber
\end{eqnarray}
\begin{eqnarray} V_{\rm noflip}={(m + \Delta/2)\over \sqrt{16\pi^3}} &&
      \sum_{s} \int d\ub{k} d\ub{l}d\ub{q}  \ths\ths{1\over l^+\sqrt{q^+}}
\nonumber \\
\times\Bigl\{ &&b^\dagger_{s,\ub{k}}b_{s,\ub{l}}a_{\ub{q}}
	\ths\kd3(\ub{k}-\ub{l}-\ub{q})
+d^\dagger_{-s,\ub{l}}b^\dagger_{s,\ub{k}}a_{\ub{q}}
	\ths\kd3(\ub{k}+\ub{l}-\ub{q}) 
\nonumber \\
+&&b^\dagger_{s,\ub{k}}b_{s,\ub{l}}a^\dagger_{\ub{q}}
	\ths\kd3(\ub{k}+\ub{q}-\ub{l})
+d^\dagger_{s,\ub{l}}d_{s,\ub{k}}a_{\ub{q}}
	\ths\kd3(\ub{k}+\ub{q}-\ub{l})
\nonumber \\ +&&d_{s,\ub{k}}b_{-s,\ub{l}}a^\dagger_{\ub{q}}
	\ths\kd3(\ub{k}+\ub{l}-\ub{q})
+d^\dagger_{s,\ub{l}}d_{s,\ub{k}}a^\dagger_{\ub{q}}
	\ths\kd3(\ub{k}-\ub{l}-\ub{q}) \Bigr\}+{\rm h.c.}  \\ 
-{\Delta/2\over \sqrt{16\pi^3}} 
  &&\sum_{s} \int d\ub{k} d\ub{l}d\ub{q}\ths\ths{1\over l^+\sqrt{q^+}}
\nonumber \\
\times\Bigl\{ &&b^\dagger_{s,\ub{k}}\b_{s,\ub{l}}a_{\ub{q}}
	\ths\kd3(\ub{k}-\ub{l}-\ub{q})
+\d^\dagger_{-s,\ub{l}}b^\dagger_{s,\ub{k}}a_{\ub{q}}
	\ths\kd3(\ub{k}+\ub{l}-\ub{q}) 
\nonumber \\
+&&b^\dagger_{s,\ub{k}}\b_{s,\ub{l}}a^\dagger_{\ub{q}}
	\ths\kd3(\ub{k}+\ub{q}-\ub{l})
+\d^\dagger_{s,\ub{l}}d_{s,\ub{k}}a_{\ub{q}}
	\ths\kd3(\ub{k}+\ub{q}-\ub{l})
\nonumber \\
+&&d_{s,\ub{k}}\b_{-s,\ub{l}}a^\dagger_{\ub{q}}
	\ths\kd3(\ub{k}+\ub{l}-\ub{q})
+\d^\dagger_{s,\ub{l}}d_{s,\ub{k}}a^\dagger_{\ub{q}}
	\ths\kd3(\ub{k}-\ub{l}-\ub{q}) \Bigr\}+{\rm h.c.} \nonumber
\end{eqnarray}
The structure of these interactions reflects the conservation of $J_z$ in
each interaction; the light-cone spin-flip $\Delta S_z = \pm 1$ of
the fermions is compensated by a unit change $\Delta L_z = \mp 1$ 
in orbital angular momentum~\cite{Brodsky:2001ii}.

Notice that no four-point
interactions arise in the light-cone Hamiltonian from the elimination of
the dependent fermionic fields.  The absence of such instantaneous
interactions follows from the lack of mass
dependence in such interactions and from the opposite signature of the PV
fermion; the interactions associated with an instantaneous ``physical''
fermion then cancel against those of the instantaneous PV fermion.

\section{Construction of the Exact Solutions}
\label{sec:Exact}

If $\delta$ and $\Delta$ are equal to zero, the system is exactly
solvable.  To see this, we define an index for each state
as the number of $\alpha^\dagger$ type quanta plus the number of
$\beta^\dagger$ and $\delta^\dagger$ type quanta minus the number of
$a^\dagger$ type quanta minus the number of $b^\dagger$ and $d^\dagger$
type quanta.  States with a definite index are then eigenstates of the 
operator ${\cal I} = [\alpha^\dagger a + \beta^\dagger b+\delta^\dagger d] 
-[a^\dagger \alpha + b^\dagger \beta +d^\dagger \delta]$ with the value
of the index being the eigenvalue; note, however, that matrix elements of
${\cal I}$ between such states may not be equal to the index, due to the
indefinite metric.  States with a definite value of the index span the
space.  The kinetic energy part of $P^-$ (for zero $\delta$ and
$\Delta$) is diagonal in ${\cal I}$, whereas the interacting part of
$P^-$, when acting on a state of given index, produces only states with
{\it lower} index.  Thus the light-cone Hamiltonian is triangular,
allowing its eigensolutions to be constructed as a combination of
Fock states in a finite number of sectors.  In particular, each 
eigenvector of the system will contain a
state of highest index, and its eigenvalue will be equal to
the free eigenvalue of the highest state.

We define the functions
\begin{eqnarray}
  U(\ub{k},\ub{l}) &\equiv& {1 \over \sqrt{16\pi^3}}
          {1 \over \sqrt{k^+ -l^+}} 
       \left( {1 \over l^+} + {1 \over k^+}\right)\,,  \\
   V(\ub{k},\ub{l}) &\equiv& {1 \over \sqrt{16\pi^3}}
                      {1 \over \sqrt{k^+ -l^+}} 
    \left( {-l_1 - i l_2 \over l^+} + {k_1 + i k_2 \over k^+}\right)\,,  \\
      T(\ub{k},\ub{l}) &\equiv& {1 \over \sqrt{16\pi^3}}
   {1 \over \sqrt{k^+}} 
   \left( {-(k_1 -l_1) + i (k_2 -l_2) \over k^+ - l^+} 
     + {l_1 - i l_2 \over l^+}\right)\,,  \\
     S(\ub{k},\ub{l}) &\equiv& {1 \over \sqrt{16\pi^3}}
         {1 \over \sqrt{k^+}} 
       \left( {-1 \over k^+ -l^+} + {1 \over l^+}\right)\,,
\end{eqnarray} 
and
\begin{equation}
E_{n,m}(\ub{l_1},\dots,\ub{l_n},\ub{l_{n+1}},\dots,\ub{l_{n+m}})
  \equiv \sum_{i = 1}^n \Bigl[ 
      {\bbox{l}_{\perp,i}^{\ths2}+m^2 \over l_i^+}\Bigr] 
    + \sum_{i = n+1}^{n+m} \Bigl[
{\bbox{l}_{\perp,i}^{\ths2}+\mu^2 \over l_i^+}\Bigr]\,.
\end{equation} 
With this notation, we can write the eigenvector whose
highest state is $\beta_{+,\ub{k}}^\dagger \ket{0}$ as
\begin{eqnarray}
    \beta_{+,\ub{k}}^\dagger \ket{0} + && mg\int_0^{k^+} d\ub{l}
{U(\ub{k},\ub{l}) \over E_{1,0}(\ub{k}) - E_{1,1}(\ub{l},\ub{k} -
\ub{l})} b_{+,\ub{l}}^\dagger a_{\ub{k} - \ub{l}}^\dagger \ket{0}
\nonumber \\ 
+ &&g \int_0^{k^+} d\ub{l} {V(\ub{k},\ub{l}) \over
E_{1,0}(\ub{k}) - E_{1,1}(\ub{l},\ub{k} - \ub{l})} b_{-,\ub{l}}^\dagger
a_{\ub{k} - \ub{l}}^\dagger \ket{0}\,.
\label{df}
\end{eqnarray} 
The eigenvalue of the state is $E_{1,0}(\ub{k})$.  Not all
states are this simple.  Another example: The eigenstate whose highest
state is 
$\beta_{+,\ub{l}}^\dagger \alpha_{\ub{k} - \ub{l}}^\dagger\ket{0}$ is
\begin{eqnarray} 
&&\beta_{+,\ub{l}}^\dagger \alpha_{\ub{k} -
\ub{l}}^\dagger \ket{0}   + mg{U(\ub{k},\ub{l}) \over
E_{1,1}(\ub{l},\ub{k} - \ub{l}) - E_{1,0}(\ub{k})} b_{+,\ub{k}}^\dagger
\ket{0} - g{V(\ub{k},\ub{l}) \over E_{1,1}(\ub{l},\ub{k} - \ub{l}) -
E_{1,0}(\ub{k})} b_{-,\ub{k}}^\dagger \ket{0}
\nonumber \\
 &&+ mg\int_0^{l^+} d\ub{t}{U(\ub{l},\ub{t}) \over E_{1,1}(\ub{l},\ub{k}
- \ub{l}) - E_{1,2}(\ub{t},\ub{l} - \ub{t},\ub{k} - \ub{l})}
b_{+,\ub{t}}^\dagger a_{\ub{l} - \ub{t}}^\dagger \alpha_{\ub{k} -
\ub{l}}^\dagger \ket{0}
\nonumber \\  
&&+ g\int_0^{l^+} d\ub{t}{V(\ub{l},\ub{t}) \over
E_{1,1}(\ub{l},\ub{k} - \ub{l}) - E_{1,2}(\ub{t},\ub{l} - \ub{t},\ub{k} -
\ub{l})} b_{-,\ub{t}}^\dagger a_{\ub{l} - \ub{t}}^\dagger \alpha_{\ub{k}
- \ub{l}}^\dagger \ket{0}
\nonumber \\
 &&+ g\int_0^{k^+ - l^+} d\ub{t}{T(\ub{k} - \ub{l},\ub{t}) \over
E_{1,1}(\ub{l},\ub{k} - \ub{l}) - E_{3,0}(\ub{t},\ub{k} - \ub{l} -
\ub{t},\ub{l})} b_{+,\ub{t}}^\dagger d_{+,\ub{k} - \ub{l} -
\ub{t}}^\dagger \beta_{+,\ub{l}}^\dagger \ket{0}
\nonumber \\  
&&- g\int_0^{k^+ - l^+} d\ub{t}{T^*(\ub{k} - \ub{l},\ub{t})
\over E_{1,1}(\ub{l},\ub{k} - \ub{l}) - E_{3,0}(\ub{t},\ub{k} - \ub{l} -
\ub{t},\ub{l})} b_{-,\ub{t}}^\dagger d_{-,\ub{k} - \ub{l} -
\ub{t}}^\dagger \beta_{+,\ub{l}}^\dagger \ket{0}
\\  
&&+ mg\int_0^{k^+ - l^+} d\ub{t}{S(\ub{k} - \ub{l},\ub{t}) \over
E_{1,1}(\ub{l},\ub{k} - \ub{l}) - E_{3,0}(\ub{t},\ub{k} - \ub{l} -
\ub{t},\ub{l})} (b_{-,\ub{t}}^\dagger d_{+,\ub{k} - \ub{l} -
\ub{t}}^\dagger + b_{+,\ub{t}}^\dagger d_{-,\ub{k} - \ub{l} -
\ub{t}}^\dagger) \beta_{+,\ub{l}}^\dagger \ket{0}
\nonumber \\
 &&+g^2\int_0^{k^+ - l^+} d\ub{t}\int_0^{l^+} d\ub{q} {T(\ub{k} -
\ub{l},\ub{t}) \over (E_{1,1}(\ub{l},\ub{k} - \ub{l}) -
E_{3,0}(\ub{t},\ub{k} - \ub{l} - \ub{t},\ub{l}))(E_{1,1}(\ub{l},\ub{k} -
\ub{l}) - E_{3,1}(\ub{t},\ub{k} - \ub{l} - \ub{t},\ub{q},\ub{l} -
\ub{q}))}
\nonumber \\ 
&& b_{+,\ub{t}}^\dagger d_{+,\ub{k} - \ub{l} -
\ub{t}}^\dagger (mU(\ub{l},\ub{q}) b_{+,\ub{q}}^\dagger a_{\ub{l} -
\ub{q}}^\dagger + V(\ub{l},\ub{q}) b_{-,\ub{q}}^\dagger a_{\ub{l} -
\ub{q}}^\dagger) \ket{0}
\nonumber \\  
&&- g^2\int_0^{k^+ - l^+} d\ub{t}\int_0^{l^+} d\ub{q}
{T^*(\ub{k} - \ub{l},\ub{t}) \over (E_{1,1}(\ub{l},\ub{k} - \ub{l}) -
E_{3,0}(\ub{t},\ub{k} - \ub{l} - \ub{t},\ub{l}))(E_{1,1}(\ub{l},\ub{k} -
\ub{l}) - E_{3,1}(\ub{t},\ub{k} - \ub{l} - \ub{t},\ub{q},\ub{l} -
\ub{q}))}
\nonumber \\ 
&& b_{-,\ub{t}}^\dagger d_{-,\ub{k} - \ub{l} -
\ub{t}}^\dagger (mU(\ub{l},\ub{q}) b_{+,\ub{q}}^\dagger a_{\ub{l} -
\ub{q}}^\dagger + V(\ub{l},\ub{q}) b_{-,\ub{q}}^\dagger a_{\ub{l} -
\ub{q}}^\dagger) \ket{0}
\nonumber \\
 &&+ mg^2\int_0^{k^+ - l^+} d\ub{t}\int_0^{l^+} d\ub{q} {S(\ub{k} -
\ub{l},\ub{t}) \over (E_{1,1}(\ub{l},\ub{k} - \ub{l}) -
E_{3,0}(\ub{t},\ub{k} - \ub{l} - \ub{t},\ub{l}))(E_{1,1}(\ub{l},\ub{k} -
\ub{l}) - E_{3,1}(\ub{t},\ub{k} - \ub{l} - \ub{t},\ub{q},\ub{l} -
\ub{q}))}
\nonumber \\ 
&&(b_{-,\ub{t}}^\dagger d_{+,\ub{k} - \ub{l} -
\ub{t}}^\dagger + b_{+,\ub{t}}^\dagger d_{-,\ub{k} - \ub{l} -
\ub{t}}^\dagger) (mU(\ub{l},\ub{q}) b_{+,\ub{q}}^\dagger a_{\ub{l} -
\ub{q}}^\dagger + V(\ub{l},\ub{q}) b_{-,\ub{q}}^\dagger a_{\ub{l} -
\ub{q}}^\dagger) \ket{0}
\nonumber  \\
 &&+g^2\int_0^{k^+ - l^+} d\ub{t}\int_0^{l^+} d\ub{q} 
{T(\ub{k} - \ub{l},\ub{t}) \over (E_{1,1}(\ub{l},\ub{k} - \ub{l}) - 
E_{1,2}(\ub{t},\ub{k} - \ub{l} - \ub{t},\ub{l}))
(E_{1,1}(\ub{l},\ub{k} - \ub{l}) - 
E_{3,1}(\ub{t},\ub{k} - \ub{l} - \ub{t},\ub{q},\ub{l} - \ub{q}))} 
\nonumber \\
&& b_{+,\ub{t}}^\dagger d_{+,\ub{k} - \ub{l} - \ub{t}}^\dagger 
(mU(\ub{l},\ub{q}) b_{+,\ub{q}}^\dagger 
a_{\ub{l} - \ub{q}}^\dagger + 
V(\ub{l},\ub{q}) b_{-,\ub{q}}^\dagger 
a_{\ub{l} - \ub{q}}^\dagger) \ket{0} 
\nonumber \\ 
&&- g^2\int_0^{k^+ - l^+} d\ub{t}\int_0^{l^+} d\ub{q} 
{T^*(\ub{k} - \ub{l},\ub{t}) \over (E_{1,1}(\ub{l},\ub{k} - \ub{l}) - 
E_{1,2}(\ub{t},\ub{k} - \ub{l} - \ub{t},\ub{l}))
(E_{1,1}(\ub{l},\ub{k} - \ub{l}) - 
E_{3,1}(\ub{t},\ub{k} - \ub{l} - \ub{t},\ub{q},\ub{l} - \ub{q}))} 
\nonumber \\
&& b_{-,\ub{t}}^\dagger d_{-,\ub{k} - \ub{l} - \ub{t}}^\dagger 
(mU(\ub{l},\ub{q}) b_{+,\ub{q}}^\dagger 
a_{\ub{l} - \ub{q}}^\dagger + 
V(\ub{l},\ub{q}) b_{-,\ub{q}}^\dagger 
a_{\ub{l} - \ub{q}}^\dagger) \ket{0} 
\nonumber \\
 &&+ mg^2\int_0^{k^+ - l^+} d\ub{t}\int_0^{l^+} d\ub{q} 
{S(\ub{k} - \ub{l},\ub{t}) \over 
(E_{1,1}(\ub{l},\ub{k} - \ub{l}) - 
E_{1,2}(\ub{t},\ub{k} - \ub{l} - \ub{t},\ub{l}))
(E_{1,1}(\ub{l},\ub{k} - \ub{l}) - 
E_{3,1}(\ub{t},\ub{k} - \ub{l} - \ub{t},\ub{q},\ub{l} - \ub{q}))} 
\nonumber \\
&&(b_{-,\ub{t}}^\dagger d_{+,\ub{k} - \ub{l} - \ub{t}}^\dagger 
+ b_{+,\ub{t}}^\dagger d_{-,\ub{k} - \ub{l} - \ub{t}}^\dagger) 
(mU(\ub{l},\ub{q}) b_{+,\ub{q}}^\dagger 
a_{\ub{l} - \ub{q}}^\dagger + V(\ub{l},\ub{q}) 
b_{-,\ub{q}}^\dagger a_{\ub{l} - \ub{q}}^\dagger) \ket{0}\,.  \nonumber
\end{eqnarray}
Notice that any state containing only $a^\dagger$'s,
$b^\dagger$'s and $d^\dagger$'s is an eigenstate of the full $P^-$ with
the same free eigenvalue.

In these examples the eigenvectors have zero norm, but this is
not always the case.  If we add the vector $b_{+,\ub{k}}^\dagger
\ket{0}$ to the state given in (\ref{df}), the resulting vector is an
eigenvector with the same eigenvalue but nonzero norm.  Two
such states with momenta $p$ and $k$ satisfy 
$\langle p|k\rangle = 2 \delta(p-k)$.
An eigenvector whose highest state (as per the index) is a state
composed purely of physical quanta corresponds to that physical state in
the free theory; all other states are unphysical. 

Even when $\delta$ or $\Delta$ is nonzero, there is interesting
structure.  The interacting part of $P^-$ is still a move-down operator
in terms of the index.  In fact the only operators which move a state up
to a higher index (and thereby fill in the upper triangle of $P^-$) are the
operators
\begin{equation}
\int d{\ub{q}}
\Bigl[ {-\delta^2\over 2q^+}\Bigr] \a^\dagger_{\ub{q}}\a_{\ub{q}}
+\sum_s \int d{\ub{k}}
\Bigl[ {-D^2\over 2k^+}\Bigr]
\Bigl(\b^\dagger_{s,\ub{k}}\b_{s,\ub{k}} +
\d^\dagger_{s,\ub{k}}\d_{s,\ub{k}}\Bigr)\,. \label{to}
\end{equation} 
These operators are still diagonal in momentum space and
are independent of the transverse momenta.

If we chose an unperturbed $P^-$ containing all the terms in the full
$P^-$ except those in (\ref{to}), we obtain a triangular system but the
operator is very deficient and probably not very useful.  On the other
hand, if we forget for the moment that $\Delta$ and $D$ are related and
treat them formally
as independent parameters, we can set $D = 0$ while
allowing
$\Delta$ to become nonzero.  If we chose as the unperturbed $P^-$, the
$P^-$ which results from setting $\delta$ and $D$ equal to zero but
allowing $\Delta$ to be nonzero, then the perturbing $P^-$ (the terms
proportional to $\delta$ and D) has no dependence on the coupling
constant, $g$.  (If $\Delta$ is a perturbation parameter, the perturbing
$P^-$ does depend on $g$ through the $V_{\rm noflip}$ interaction.) The
$P^-$ which includes $\Delta$, but not $\delta$ or $D$, is not deficient
and may form a good starting point for calculations.  In that case the
eigenvectors project onto an infinite number of sectors.  For instance,
the eigenvector whose highest state is $\beta_{+,\ub{k}}^\dagger \ket{0}$
projects onto all sectors containing a $\beta_+^\dagger$ or a $b^\dagger$
particle plus an arbitrary number of $a^\dagger$ particles.  We can write
the eigenvector as
\begin{eqnarray}
    \beta_{+,\ub{k}}^\dagger \ket{0} + &&\sum_{i = 1}^\infty \int_0^{k^+}
d\ub{l}^+_1 \int_0^{{l}_1^+} d\ub{l}^+_2 \dots \int_0^{{l}_{i-1}^+}
d\ub{l}^+_i \nonumber \\ && \left[X_i(\ub{k},\ub{l}_1,
\dots,\ub{l}_i)\beta_{+,\ub{l}_i}^\dagger a_{\ub{k} - \ub{l}_1}^\dagger
a_{\ub{l}_1 - \ub{l}_2}^\dagger \dots a_{\ub{l}_{i-1} - \ub{l}_i}^\dagger
\ket{0} \right.\nonumber \\ + &&Y_i(\ub{k},\ub{l}_1, \dots,\ub{l}_i)
b_{+,\ub{l}_i}^\dagger a_{\ub{k} - \ub{l}_1}^\dagger a_{\ub{l}_1 -
\ub{l}_2}^\dagger \dots a_{\ub{l}_{i-1} - \ub{l}_i}^\dagger \ket{0}
\nonumber \\ + &&\left.Z_i(\ub{k},\ub{l}_1, \dots,\ub{l}_i)
b_{-,\ub{l}_i}^\dagger a_{\ub{k} - \ub{l}_1}^\dagger a_{\ub{l}_1 -
\ub{l}_2}^\dagger \dots a_{\ub{l}_{i-1} - \ub{l}_i}^\dagger
\ket{0}\right]\,.
\end{eqnarray} 
The $X$, $Y$, and $Z$'s satisfy the following recursion relations:
\begin{eqnarray}
      X_{i+1}(\ub{k},\ub{l}_1, \dots,\ub{l}_{i+1}) &=& 
         {- \ha \Delta g U(\ub{l}_{i},\ub{l}_{i+1})
    X_i(\ub{k},\ub{l}_1, \dots,\ub{l}_{i}) \over E_{1,0}(\ub{k}) -
E_{1,i+1}(\ub{l}_{i+1},\ub{k} - \ub{l}_1,\ub{l}_1 - \ub{l}_2,
\dots,\ub{l}_i - \ub{l}_{i+1})} \,,
\\
      Y_{i+1}(\ub{k},\ub{l}_1, \dots,\ub{l}_{i+1}) &=& {- \ha \Delta g
U(\ub{l}_{i},\ub{l}_{i+1})Y_{i}(\ub{k},\ub{l}_1, \dots,\ub{l}_{i}) + (m +
\ha \Delta) g U(\ub{l}_{i},\ub{l}_{i+1})X_{i}(\ub{k},\ub{l}_1,
\dots,\ub{l}_{i}) \over E_{1,0}(\ub{k}) - E_{1,i+1}(\ub{l}_{i+1},\ub{k} -
\ub{l}_1,\ub{l}_1 - \ub{l}_2, \dots,\ub{l}_i - \ub{l}_{i+1})}\,,
\\
      Z_{i+1}(\ub{k},\ub{l}_1, \dots,\ub{l}_{i+1}) &=& {- \ha \Delta g
U(\ub{l}_{i},\ub{l}_{i+1})Z_{i}(\ub{k},\ub{l}_1, \dots,\ub{l}_{i}) +  g
V(\ub{l}_{i},\ub{l}_{i+1})X_{i}(\ub{k},\ub{l}_1, \dots,\ub{l}_{i}) \over
E_{1,0}(\ub{k}) - E_{1,i+1}(\ub{l}_{i+1},\ub{k} - \ub{l}_1,\ub{l}_1 -
\ub{l}_2, \dots,\ub{l}_i - \ub{l}_{i+1})} \,.
\end{eqnarray} 
These are subject to the initial conditions
\begin{eqnarray} 
X_1(\ub{k},\ub{l}_1) &=& - \ha g \Delta
U(\ub{k},\ub{l}_1)\,, 
\\
Y_1(\ub{k},\ub{l}_1) &=& (m + \ha \Delta) g U(\ub{k},\ub{l}_1)\,, 
\\
Z_1(\ub{k},\ub{l}_1) &=&  g  V(\ub{k},\ub{l}_1)\,.
\end{eqnarray}

\section{The Operator Solution, the Vacuum and the Equal-Time
Representation}
\label{sec:OpSoln}

We now give the operator solution for the case $\delta = \Delta = D =0$.
The simplest field to obtain is the Bose field.  If we define the
operator, $A_{\ub{q}}$, to be
\begin{eqnarray}
      &&A_{\ub{q}} \equiv \alpha_{\ub{q}} \nonumber \\
 +&& g\int_{q^+}^\infty d\ub{k}  {1 \over E_{0,1}(\ub{q})+E_{1,0}(\ub{k}
- \ub{q}) - E_{1,0}(\ub{k})} (V(\ub{k}, \ub{k} - \ub{q}) b_{-,\ub{k} -
\ub{q}}^\dagger b_{+,\ub{k}} - V^*(\ub{k}, \ub{k} - \ub{q}) b_{+, \ub{k}
- \ub{q}}^\dagger b_{-,\ub{k}})
\nonumber \\ 
+&& g\int_0^{q^+} d\ub{k} {1 \over
E_{0,1}(\ub{q})-E_{1,0}(\ub{q} - \ub{k}) - E_{1,0}(\ub{k})} (T^*(\ub{q},
\ub{k}) d_{+,\ub{q} - \ub{k}} b_{+,\ub{k}} - T(\ub{q}, \ub{k}) d_{-,
\ub{q} - \ub{k}} b_{-,\ub{k}})
\nonumber \\ 
+&& g\int_{q^+}^\infty d\ub{k}{1 \over
E_{0,1}(\ub{q})+E_{1,0}(\ub{k} - \ub{q}) - E_{1,0}(\ub{k})} (V(\ub{k},
\ub{k} - \ub{q}) d_{-,\ub{k} - \ub{q}}^\dagger d_{+,\ub{k}} - V^*(\ub{k},
\ub{k} - \ub{q}) d_{+, \ub{k} - \ub{q}}^\dagger d_{-,\ub{k}})
\nonumber \\ 
+&& mg\int_{q^+}^\infty d\ub{k}{U(\ub{k}, \ub{k} - \ub{q})
\over E_{0,1}(\ub{q})+E_{1,0}(\ub{k} - \ub{q}) - E_{1,0}(\ub{k})} (
b_{+,\ub{k} - \ub{q}}^\dagger b_{+,\ub{k}} +  b_{-, \ub{k} -
\ub{q}}^\dagger b_{-,\ub{k}})
\nonumber \\ 
+&& mg\int_0^{q^+} d\ub{k}{S(\ub{q}, \ub{k}) \over
E_{0,1}(\ub{q})-E_{1,0}(\ub{q} - \ub{k}) - E_{1,0}(\ub{k})} ( d_{+,\ub{q}
- \ub{k}} b_{-,\ub{k}} +  d_{-, \ub{q} - \ub{k}} b_{+,\ub{k}})
\nonumber \\ 
+&& mg\int_{q^+}^\infty d\ub{k}{U(\ub{k}, \ub{k} - \ub{q})
\over E_{0,1}(\ub{q})+E_{1,0}(\ub{k} - \ub{q}) - E_{1,0}(\ub{k})} (
d_{+,\ub{k} - \ub{q}}^\dagger d_{+,\ub{k}} +  d_{-, \ub{k} -
\ub{q}}^\dagger d_{-,\ub{k}})\,,
\end{eqnarray} 
we can use the relation
\begin{equation}
        [P^- , \zeta] = -i\partial^-\zeta\,,
\end{equation} 
along with the initial condition (\ref{ichi}), to show that
\begin{eqnarray}
    &&\zeta(x^+,\ub{x}) = {1\over\sqrt{16\pi^3}} \int d\ub{q}
{1\over\sqrt{q^+}} A_{\ub{q}} e^{-i(\ha E_{0,1}(\ub{q})x^+ +
\ub{q}\cdot\ub{x})}  - {1\over\sqrt{16\pi^3}} \int d\ub{q}
{1\over\sqrt{q^+}}
 e^{-i\ub{q}\cdot\ub{x}} \Biggl[
\nonumber \\ &&g\int_{q^+}^\infty d\ub{k} {e^{i \ha (E_{1,0}(\ub{k} -
\ub{q}) - E_{1,0}(\ub{k}))x^+} \over E_{0,1}(\ub{q})+E_{1,0}(\ub{k} -
\ub{q}) - E_{1,0}(\ub{k})}  (V(\ub{k}, \ub{k} - \ub{q}) b_{-,\ub{k} -
\ub{q}}^\dagger b_{+,\ub{k}} - V^*(\ub{k}, \ub{k} - \ub{q}) b_{+, \ub{k}
- \ub{q}}^\dagger b_{-,\ub{k}})
\nonumber \\
 +&&g\int_0^{q^+}d\ub{k} {e^{i \ha (-E_{1,0}(\ub{q} - \ub{k}) -
E_{1,0}(\ub{k}))x^+} \over E_{0,1}(\ub{q})-E_{1,0}(\ub{q} - \ub{k}) -
E_{1,0}(\ub{k})}  (T^*(\ub{q}, \ub{k}) d_{+,\ub{q} - \ub{k}} b_{+,\ub{k}}
- T(\ub{q}, \ub{k}) d_{-, \ub{q} - \ub{k}} b_{-,\ub{k}})
\nonumber \\
 +&&g\int_{q^+}^\infty d\ub{k}{e^{i \ha (E_{1,0}(\ub{k} - \ub{q}) -
E_{1,0}(\ub{k}))x^+} \over E_{0,1}(\ub{q})+E_{1,0}(\ub{k} - \ub{q}) -
E_{1,0}(\ub{k})}  (V(\ub{k}, \ub{k} - \ub{q}) d_{-,\ub{k} -
\ub{q}}^\dagger d_{+,\ub{k}} - V^*(\ub{k}, \ub{k} - \ub{q}) d_{+, \ub{k}
- \ub{q}}^\dagger d_{-,\ub{k}})
\nonumber \\
 +&& mg\int_{q^+}^\infty d\ub{k}{e^{i \ha (E_{1,0}(\ub{k} - \ub{q}) -
E_{1,0}(\ub{k}))x^+}U(\ub{k}, \ub{k} - \ub{q}) \over
E_{0,1}(\ub{q})+E_{1,0}(\ub{k} - \ub{q}) - E_{1,0}(\ub{k})}  (
b_{+,\ub{k} - \ub{q}}^\dagger b_{+,\ub{k}} + b_{-, \ub{k} -
\ub{q}}^\dagger b_{-,\ub{k}})
\nonumber \\
 +&& mg\int_0^{q^+}d\ub{k} {e^{i \ha (-E_{1,0}(\ub{q} - \ub{k}) -
E_{1,0}(\ub{k}))x^+}S(\ub{q}, \ub{k}) \over
E_{0,1}(\ub{q})-E_{1,0}(\ub{q} - \ub{k}) - E_{1,0}(\ub{k})}  (
d_{+,\ub{q} - \ub{k}} b_{-,\ub{k}} +  d_{-, \ub{q} - \ub{k}}
b_{+,\ub{k}})
\nonumber \\
 +&& mg\int_{q^+}^\infty d\ub{k} {e^{i \ha (E_{1,0}(\ub{k} - \ub{q}) -
E_{1,0}(\ub{k}))x^+}U(\ub{k}, \ub{k} - \ub{q}) \over
E_{0,1}(\ub{q})+E_{1,0}(\ub{k} - \ub{q}) - E_{1,0}(\ub{k})}  (
d_{+,\ub{k} - \ub{q}}^\dagger d_{+,\ub{k}} +  d_{-, \ub{k} -
\ub{q}}^\dagger d_{-,\ub{k}}) \Biggr] + {\rm h.c.}
\end{eqnarray} 
By evaluating this expression at $t = 0$ (or any other
equal-time surface) we can work out the relation between the light-cone
operators and the equal-time operators.  We shall indicate three-vectors
in the three spatial dimensions with a hat: $\hat k \equiv
(k_1,k_2,k_3)$, and will indicate the equal-time operators with a breve:
$\breve a$.  We write as usual
\begin{equation}
   \zeta(0,\hat x ) = {1 \over \sqrt{8\pi^3}} \int d\hat k {1 \over
\sqrt{2 \omega_{\hat k}}}\left( \breve a_{\hat k} e^{i \hat k \cdot \hat
x} + \breve a_{\hat k}^\dagger e^{-i \hat k \cdot \hat x} \right),
\end{equation} 
where
\begin{equation}
   \breve a_{\hat k} = {1 \over \sqrt{8\pi^3}}{1 \over \sqrt{2}} \int
d\hat x e^{-i \hat k \cdot \hat x}\sqrt{\omega_{\hat k}} \left(
\zeta(0,\hat x) + {i \over \omega_{\hat k}} \partial_0 \zeta(0,\hat x)
\right).
\end{equation} 
In these and later formulas we use 
$\omega_{\hat k} \equiv \sqrt{\mu^2 + \hat k^2}$.  
It is also useful to define the quantities
\begin{equation}
    p^-(\ub{q},\ub{k}) \equiv -E_{1,0}(\ub{k} - \ub{q}) +
E_{1,0}(\ub{k})\,,
  \quad \quad
    r^-(\ub{q},\ub{k}) \equiv E_{1,0}(\ub{q} - \ub{k}) + E_{1,0}(\ub{k}),
\end{equation} 
with which we can define
\begin{eqnarray}
     && p_0(\ub{q},\ub{k}) \equiv \ha(p^-(\ub{q},\ub{k}) + q^+)\,, \quad
\quad p_3(\ub{q},\ub{k}) \equiv \ha(p^-(\ub{q},\ub{k}) - q^+)\,,
\\
     && r_0(\ub{q},\ub{k}) \equiv \ha(r^-(\ub{q},\ub{k}) + q^+)\,, \quad
\quad r_3(\ub{q},\ub{k}) \equiv \ha(r^-(\ub{q},\ub{k}) - q^+)\,.
\end{eqnarray} 
These allow us to define the spatial three-vectors
\begin{equation}
      \hat p(\ub{q},\ub{k}) \equiv (\senk{q},p_3)\,, \quad  \quad
       \hat r(\ub{q},\ub{k}) \equiv (\senk{q},r_3)\,.
\end{equation} 
We now find
\begin{eqnarray}
    &&\breve a_{\hat t} = \sqrt{{\omega_{\hat t} - t_3 \over
\omega_{\hat t}}} A_{\omega_{\hat t} - t_3,\senk{t}}   -  \int d\ub{q}
{1 \over 2\sqrt{q^+}} \Biggl[
\nonumber \\ 
&&g\int_{q^+}^\infty d\ub{k} { (\sqrt{\omega_{\hat t}} +
{p_0(\ub{q},\ub{k}) \over \sqrt{\omega_{\hat t}}}) \delta(\hat t - \hat
p(\ub{q},\ub{k})) \over E_{0,1}(\ub{q})+E_{1,0}(\ub{k} - \ub{q}) -
E_{1,0}(\ub{k})}  (V(\ub{k}, \ub{k} - \ub{q}) b_{-,\ub{k} -
\ub{q}}^\dagger b_{+,\ub{k}} - V^*(\ub{k}, \ub{k} - \ub{q}) b_{+, \ub{k}
- \ub{q}}^\dagger b_{-,\ub{k}})
\nonumber \\ 
+&&g\int_{q^+}^\infty d\ub{k} { (\sqrt{\omega_{\hat t}} -
{p_0(\ub{q},\ub{k}) \over \sqrt{\omega_{\hat t}}}) \delta(\hat t + \hat
p(\ub{q},\ub{k})) \over E_{0,1}(\ub{q})+E_{1,0}(\ub{k} - \ub{q}) -
E_{1,0}(\ub{k})}  (V^*(\ub{k}, \ub{k} - \ub{q})  b_{+,\ub{k}}^\dagger
b_{-,\ub{k} - \ub{q}} - V(\ub{k}, \ub{k} - \ub{q})  b_{-,\ub{k}}^\dagger
b_{+, \ub{k} - \ub{q}})
\nonumber \\
 +&&g\int_0^{q^+} d\ub{k} {(\sqrt{\omega_{\hat t}} + {r_0(\ub{q},\ub{k})
\over \sqrt{\omega_{\hat t}}}) \delta(\hat t - \hat r(\ub{q},\ub{k}))
\over E_{0,1}(\ub{q})-E_{1,0}(\ub{q} - \ub{k}) - E_{1,0}(\ub{k})}
(T^*(\ub{q}, \ub{k}) d_{+,\ub{q} - \ub{k}} b_{+,\ub{k}} - T(\ub{q},
\ub{k}) d_{-, \ub{q} - \ub{k}} b_{-,\ub{k}})
\nonumber \\ 
+&&g\int_0^{q^+} d\ub{k} {(\sqrt{\omega_{\hat t}} -
{r_0(\ub{q},\ub{k}) \over \sqrt{\omega_{\hat t}}}) \delta(\hat t + \hat
r(\ub{q},\ub{k})) \over E_{0,1}(\ub{q})-E_{1,0}(\ub{q} - \ub{k}) -
E_{1,0}(\ub{k})}  (T(\ub{q}, \ub{k})  b_{+,\ub{k}}^\dagger d_{+,\ub{q} -
\ub{k}}^\dagger - T^*(\ub{q}, \ub{k})  b_{-,\ub{k}}^\dagger d_{-, \ub{q}
- \ub{k}}^\dagger )
\nonumber \\
 +&&g\int_{q^+}^\infty d\ub{k}{(\sqrt{\omega_{\hat t}} +
{p_0(\ub{q},\ub{k}) \over \sqrt{\omega_{\hat t}}}) \delta(\hat t - \hat
p(\ub{q},\ub{k})) \over E_{0,1}(\ub{q})+E_{1,0}(\ub{k} - \ub{q}) -
E_{1,0}(\ub{k})}  (V(\ub{k}, \ub{k} - \ub{q}) d_{-,\ub{k} -
\ub{q}}^\dagger d_{+,\ub{k}} - V^*(\ub{k}, \ub{k} - \ub{q}) d_{+, \ub{k}
- \ub{q}}^\dagger d_{-,\ub{k}})
\nonumber \\ 
+&&g\int_{q^+}^\infty d\ub{k}{(\sqrt{\omega_{\hat t}} -
{p_0(\ub{q},\ub{k}) \over \sqrt{\omega_{\hat t}}}) \delta(\hat t + \hat
p(\ub{q},\ub{k})) \over E_{0,1}(\ub{q})+E_{1,0}(\ub{k} - \ub{q}) -
E_{1,0}(\ub{k})}  (V^*(\ub{k}, \ub{k} - \ub{q})  d_{+,\ub{k}}^\dagger
d_{-,\ub{k} - \ub{q}} - V(\ub{k}, \ub{k} - \ub{q})  d_{-,\ub{k}}^\dagger
d_{+, \ub{k} - \ub{q}})
\nonumber \\
 +&& mg\int_{q^+}^\infty d\ub{k}{(\sqrt{\omega_{\hat t}} +
{p_0(\ub{q},\ub{k}) \over \sqrt{\omega_{\hat t}}}) \delta(\hat t - \hat
p(\ub{q},\ub{k}))U(\ub{k}, \ub{k} - \ub{q}) \over
E_{0,1}(\ub{q})+E_{1,0}(\ub{k} - \ub{q}) - E_{1,0}(\ub{k})}  (
b_{+,\ub{k} - \ub{q}}^\dagger b_{+,\ub{k}} +  b_{-, \ub{k} -
\ub{q}}^\dagger b_{-,\ub{k}})   
\nonumber \\ 
+&& mg\int_{q^+}^\infty
d\ub{k}{(\sqrt{\omega_{\hat t}} - {p_0(\ub{q},\ub{k}) \over
\sqrt{\omega_{\hat t}}}) \delta(\hat t + \hat p(\ub{q},\ub{k}))U(\ub{k},
\ub{k} - \ub{q}) \over E_{0,1}(\ub{q})+E_{1,0}(\ub{k} - \ub{q}) -
E_{1,0}(\ub{k})}  (  b_{+,\ub{k}}^\dagger b_{+,\ub{k} - \ub{q}} +
b_{-,\ub{k}}^\dagger b_{-, \ub{k} - \ub{q}} )
\nonumber \\
 +&& mg\int_0^{q^+} d\ub{k} {(\sqrt{\omega_{\hat t}} +
{r_0(\ub{q},\ub{k}) \over \sqrt{\omega_{\hat t}}}) \delta(\hat t - \hat
r(\ub{q},\ub{k}))S(\ub{q}, \ub{k}) \over E_{0,1}(\ub{q})-E_{1,0}(\ub{q} -
\ub{k}) - E_{1,0}(\ub{k})}  ( d_{+,\ub{q} - \ub{k}} b_{-,\ub{k}} +  d_{-,
\ub{q} - \ub{k}} b_{+,\ub{k}})
\nonumber \\ 
+&& mg\int_0^{q^+} d\ub{k} {(\sqrt{\omega_{\hat t}} -
{r_0(\ub{q},\ub{k}) \over \sqrt{\omega_{\hat t}}}) \delta(\hat t + \hat
r(\ub{q},\ub{k}))S(\ub{q}, \ub{k}) \over E_{0,1}(\ub{q})-E_{1,0}(\ub{q} -
\ub{k}) - E_{1,0}(\ub{k})}  (  b_{-,\ub{k}}^\dagger d_{+,\ub{q} -
\ub{k}}^\dagger +   b_{+,\ub{k}}^\dagger d_{-, \ub{q} - \ub{k}}^\dagger)  
\nonumber \\
 +&& mg\int_{q^+}^\infty d\ub{k} {(\sqrt{\omega_{\hat t}} +
{p_0(\ub{q},\ub{k}) \over \sqrt{\omega_{\hat t}}}) \delta(\hat t - \hat
p(\ub{q},\ub{k}))U(\ub{k}, \ub{k} - \ub{q}) \over
E_{0,1}(\ub{q})+E_{1,0}(\ub{k} - \ub{q}) - E_{1,0}(\ub{k})}  (
d_{+,\ub{k} - \ub{q}}^\dagger d_{+,\ub{k}} +  d_{-, \ub{k} -
\ub{q}}^\dagger d_{-,\ub{k}})
\nonumber \\ 
+&& mg\int_{q^+}^\infty d\ub{k} {(\sqrt{\omega_{\hat t}} -
{p_0(\ub{q},\ub{k}) \over \sqrt{\omega_{\hat t}}}) \delta(\hat t + \hat
p(\ub{q},\ub{k}))U(\ub{k}, \ub{k} - \ub{q}) \over
E_{0,1}(\ub{q})+E_{1,0}(\ub{k} - \ub{q}) - E_{1,0}(\ub{k})}  (
d_{+,\ub{k}}^\dagger d_{+,\ub{k} - \ub{q}} +   d_{-,\ub{k}}^\dagger d_{-,
\ub{k} - \ub{q}} )\Biggr]\,.
\label{achk}
\end{eqnarray}

The relationship between
the light-cone representation and the equal-time representation is quite
complicated even for this relatively simple case.  We also see
that the equal-time perturbative vacuum is not the physical
vacuum.  The physical vacuum --- the ground state of the system, which we
shall call $\ket{\Omega}$ --- is equal to the light-cone perturbative
vacuum.  That is, the state we have called $\ket{0}$, which is destroyed
by all the light-cone destruction operators, is the physical vacuum:
$\ket{\Omega} = \ket{0}$.  We shall call the equal-time perturbative
vacuum --- the state destroyed by all the equal-time destruction
operators --- $\ket{\breve 0}$.  From the presence in (\ref{achk}) of the
terms proportional to $b_{s,\ub{k}}^\dagger d_{-s,\ub{q} -
\ub{k}}^\dagger$ and $b_{s,\ub{k}}^\dagger d_{s,\ub{q} - \ub{k}}^\dagger$
we see that $\ket{\breve 0} \not= \ket{\Omega}$.  Of course, this
conclusion immediately follows from the fact that the Hamiltonian contains
terms proportional to $\breve {b}_{s,\hat k}^\dagger \breve {d}_{t,\hat
q}^\dagger \breve {a}_{-\hat k - \hat q}^\dagger$ --- the usual
way of seeing that $\ket{\breve 0} \not= \ket{\Omega}$.

The same procedures which we have used to find the eigenstates in the
light-cone representation can be used in the equal-time representation.
However, the eigensolutions are much more complicated when expressed in
the equal-time representation than when expressed in the light-cone
representation.  For instance, although $\ket{\Omega}$ is just given by
$\ket{0}$,
$\ket{\Omega}$ projects onto any state containing N equal-time
fermion-anti fermion pairs along with N equal-time boson quanta as long as
the total momentum is zero.  Similarly, the state given in (\ref{df})
projects onto an infinite number of sectors of equal-time basis states.

We return to working out the operator solution.  We define
\begin{eqnarray}
      &&B_{s,\ub{k}} \equiv \beta_{s,\ub{k}}
\nonumber \\
 +&& {1 \over \sqrt{8\pi^3}} g\int_{k^+}^\infty d\ub{l}  {1 \over
E_{1,0}(\ub{k}) - E_{1,0}(\ub{l}) + E_{0,1}(\ub{l} - \ub{k})} \left(
{{\bbox \epsilon}_{\perp,-2s}\cdot \senk{k} \over k^+ \sqrt{l^+ - k^+}} +
{{\bbox \epsilon}_{\perp,2s}^* \cdot \senk{l} \over l^+ \sqrt{l^+ - k^+}}
\right) b_{-s,\ub{l}} a_{\ub{l} - \ub{k}}^\dagger
\nonumber \\ 
+&&{1 \over \sqrt{8\pi^3}} g\int_0^{k^+} d\ub{l}  {1 \over
E_{1,0}(\ub{k}) - E_{1,0}(\ub{l}) - E_{0,1}(\ub{k} - \ub{l})} \left(
{{\bbox \epsilon}_{\perp,-2s}\cdot \senk{k} \over k^+ \sqrt{k^+ - l^+}} +
{{\bbox \epsilon}_{\perp,2s}^* \cdot \senk{l} \over l^+ \sqrt{k^+ - l^+}}
\right) b_{-s,\ub{l}} a_{\ub{k} - \ub{l}}
\nonumber \\ 
+&&{1 \over \sqrt{8\pi^3}} g\int d\ub{l}  {1 \over
E_{1,0}(\ub{k}) + E_{1,0}(\ub{l}) - E_{0,1}(\ub{l} + \ub{k})} \left(
{{\bbox \epsilon}_{\perp,-2s}\cdot \senk{k} \over k^+ \sqrt{k^+ + l^+}} +
{{\bbox \epsilon}_{\perp,2s}^* \cdot \senk{l} \over l^+ \sqrt{k^+ + l^+}}
\right) d_{s,\ub{l}}^\dagger a_{\ub{l} + \ub{k}}
\\ +&&{1 \over \sqrt{16\pi^3}} mg\int_{k^+}^\infty d\ub{l}  {1 \over
E_{1,0}(\ub{k}) - E_{1,0}(\ub{l}) + E_{0,1}(\ub{l} - \ub{k})} \left( {1
\over k^+ \sqrt{l^+ - k^+}} + {1 \over l^+ \sqrt{l^+ - k^+}} \right)
b_{s,\ub{l}} a_{\ub{l} - \ub{k}}^\dagger    
\nonumber \\ 
+&&{1 \over
\sqrt{16\pi^3}} mg\int_0^{k^+} d\ub{l}  {1 \over E_{1,0}(\ub{k}) -
E_{1,0}(\ub{l}) - E_{0,1}(\ub{k} - \ub{l})} \left( {1 \over k^+ \sqrt{k^+
- l^+}} + {1 \over l^+ \sqrt{k^+ - l^+}} \right) b_{s,\ub{l}} a_{\ub{k} -
\ub{l}}
\nonumber \\ +&&{1 \over \sqrt{16\pi^3}} mg\int d\ub{l}  {1 \over
E_{1,0}(\ub{k}) + E_{1,0}(\ub{l}) - E_{0,1}(\ub{l} + \ub{k})} \left( {1
\over k^+ \sqrt{k^+ + l^+}} - {1 \over l^+ \sqrt{k^+ + l^+}} \right)
d_{-s,\ub{l}}^\dagger a_{\ub{l} + \ub{k}}\,,
\nonumber
\end{eqnarray}
and
\begin{eqnarray}
   &&D_{s,\ub{k}}^\dagger \equiv \delta_{s,\ub{k}}^\dagger
\nonumber \\
 +&& {1 \over \sqrt{8\pi^3}} g\int_{k^+}^\infty d\ub{l}  {1 \over
E_{1,0}(\ub{k}) - E_{1,0}(\ub{l}) + E_{0,1}(\ub{l} - \ub{k})} \left(
{-{\bbox \epsilon}_{\perp,2s}\cdot \senk{k} \over k^+ \sqrt{l^+ - k^+}} +
{-{\bbox \epsilon}_{\perp,-2s}^* \cdot \senk{l} \over l^+ \sqrt{l^+ -
k^+}} \right) d_{-s,\ub{l}}^\dagger a_{\ub{l} - \ub{k}}    
\nonumber \\
+&&{1 \over \sqrt{8\pi^3}} g\int_0^{k^+} d\ub{l}  {1 \over
E_{1,0}(\ub{k}) - E_{1,0}(\ub{l}) - E_{0,1}(\ub{k} - \ub{l})} \left(
{-{\bbox \epsilon}_{\perp,2s}\cdot \senk{k} \over k^+ \sqrt{k^+ - l^+}} +
{-{\bbox \epsilon}_{\perp,-2s}^* \cdot \senk{l} \over l^+ \sqrt{k^+ -
l^+}} \right) d_{-s,\ub{l}}^\dagger a_{\ub{k} - \ub{l}}^\dagger
\nonumber \\ 
+&&{1 \over \sqrt{8\pi^3}} g\int d\ub{l}  {1 \over
E_{1,0}(\ub{k}) + E_{1,0}(\ub{l}) - E_{0,1}(\ub{l} + \ub{k})} \left(
{-{\bbox \epsilon}_{\perp,2s}\cdot \senk{k} \over k^+ \sqrt{k^+ + l^+}} +
{-{\bbox \epsilon}_{\perp,-2s}^* \cdot \senk{l} \over l^+ \sqrt{k^+ +
l^+}} \right) b_{s,\ub{l}} a_{\ub{l} + \ub{k}}^\dagger
\\ +&&{1 \over \sqrt{16\pi^3}} mg\int_{k^+}^\infty d\ub{l}  {1 \over
E_{1,0}(\ub{k}) - E_{1,0}(\ub{l}) + E_{0,1}(\ub{l} - \ub{k})} \left( {1
\over k^+ \sqrt{l^+ - k^+}} + {1 \over l^+ \sqrt{l^+ - k^+}} \right)
d_{s,\ub{l}}^\dagger a_{\ub{l} - \ub{k}}
\nonumber \\ 
+&&{1 \over \sqrt{16\pi^3}} mg\int_0^{k^+} d\ub{l}  {1 \over
E_{1,0}(\ub{k}) - E_{1,0}(\ub{l}) - E_{0,1}(\ub{k} - \ub{l})} \left( {1
\over k^+ \sqrt{k^+ - l^+}} + {1 \over l^+ \sqrt{k^+ - l^+}} \right)
d_{s,\ub{l}}^\dagger a_{\ub{k} - \ub{l}}^\dagger    
\nonumber \\ 
+&&{1\over \sqrt{16\pi^3}} mg\int d\ub{l}  {1 \over E_{1,0}(\ub{k}) +
E_{1,0}(\ub{l}) - E_{0,1}(\ub{l} + \ub{k})} \left( {1 \over k^+ \sqrt{k^+
+ l^+}} - {1 \over l^+ \sqrt{k^+ + l^+}} \right) b_{-s,\ub{l}} a_{\ub{l}
+ \ub{k}}^\dagger.
\nonumber
\end{eqnarray} 
With these definitions we can write the full space-time
dependence of the $\eta_+$ field as
\begin{eqnarray} 
&&\eta_+(x^+,\ub{x})={1\over\sqrt{16\pi^3}} \sum_s \int
d\ub{k} \chi_s \Biggl\{e^{-i\ub{k}\cdot\ub{x}} \Biggl[ B_{s,\ub{k}} e^{-i
\ha E_{1,0}(\ub{k}) x^+}
\nonumber \\
 -&& {1 \over \sqrt{8\pi^3}} g\int_{k^+}^\infty d\ub{l}  {e^{i \ha
(-E_{1,0}(\ub{l}) + E_{0,1}(\ub{l} - \ub{k}))x^+}  \over E_{1,0}(\ub{k})
- E_{1,0}(\ub{l}) + E_{0,1}(\ub{l} - \ub{k})} \left( {{\bbox
\epsilon}_{\perp,-2s}\cdot \senk{k} \over k^+ \sqrt{l^+ - k^+}} + {{\bbox
\epsilon}_{\perp,2s}^* \cdot \senk{l} \over l^+ \sqrt{l^+ - k^+}} \right)
b_{-s,\ub{l}} a_{\ub{l} - \ub{k}}^\dagger
\nonumber \\ 
-&&{1 \over \sqrt{8\pi^3}} g\int_0^{k^+} d\ub{l}  {e^{i
\ha(- E_{1,0}(\ub{l}) - E_{0,1}(\ub{k} - \ub{l}))x^+} \over
E_{1,0}(\ub{k}) - E_{1,0}(\ub{l}) - E_{0,1}(\ub{k} - \ub{l})} \left(
{{\bbox \epsilon}_{\perp,-2s}\cdot \senk{k} \over k^+ \sqrt{k^+ - l^+}} +
{{\bbox \epsilon}_{\perp,2s}^* \cdot \senk{l}\over l^+ \sqrt{k^+ - l^+}}
\right) b_{-s,\ub{l}} a_{\ub{k} - \ub{l}}
\nonumber \\ 
-&&{1 \over \sqrt{8\pi^3}} g\int d\ub{l}  {e^{i
\ha(E_{1,0}(\ub{l}) - E_{0,1}(\ub{l} + \ub{k}))x^+} \over E_{1,0}(\ub{k})
+ E_{1,0}(\ub{l}) - E_{0,1}(\ub{l} + \ub{k})} \left( {{\bbox
\epsilon}_{\perp,-2s}\cdot \senk{k}\over k^+ \sqrt{k^+ + l^+}} + {{\bbox
\epsilon}_{\perp,2s}^* \cdot \senk{l}\over l^+ \sqrt{k^+ + l^+}} \right)
d_{s,\ub{l}}^\dagger a_{\ub{l} + \ub{k}}
\nonumber \\ 
-&&{1 \over \sqrt{16\pi^3}} mg\int_{k^+}^\infty d\ub{l}
{e^{i \ha(- E_{1,0}(\ub{l}) + E_{0,1}(\ub{l} - \ub{k}))x^+} \over
E_{1,0}(\ub{k}) - E_{1,0}(\ub{l}) + E_{0,1}(\ub{l} - \ub{k})} \left( {1
\over k^+ \sqrt{l^+ - k^+}} + {1 \over l^+ \sqrt{l^+ - k^+}} \right)
b_{s,\ub{l}} a_{\ub{l} - \ub{k}}^\dagger
\nonumber \\ 
-&&{1 \over \sqrt{16\pi^3}} mg\int_0^{k^+} d\ub{l}  {e^{i
\ha(- E_{1,0}(\ub{l}) - E_{0,1}(\ub{k} - \ub{l}))x^+} \over
E_{1,0}(\ub{k}) - E_{1,0}(\ub{l}) - E_{0,1}(\ub{k} - \ub{l})} \left( {1
\over k^+ \sqrt{k^+ - l^+}} + {1 \over l^+ \sqrt{k^+ - l^+}} \right)
b_{s,\ub{l}} a_{\ub{k} - \ub{l}}
\nonumber \\ 
-&&{1 \over \sqrt{16\pi^3}} mg\int d\ub{l}  {e^{i
\ha(E_{1,0}(\ub{l}) - E_{0,1}(\ub{l} + \ub{k}))x^+} \over E_{1,0}(\ub{k})
+ E_{1,0}(\ub{l}) - E_{0,1}(\ub{l} + \ub{k})} \left( {1 \over k^+
\sqrt{k^+ + l^+}} - {1 \over l^+ \sqrt{k^+ + l^+}} \right)
d_{-s,\ub{l}}^\dagger a_{\ub{l} + \ub{k}} \Biggr]
\nonumber \\ 
&&+ e^{+i\ub{k}\cdot\ub{x}} \Biggl[D^\dagger_{-s,\ub{k}}
e^{i \ha E_{1,0}(\ub{k})x^+}
\\ -&& {1 \over \sqrt{8\pi^3}} g\int_{k^+}^\infty d\ub{l}  {e^{i \ha
(E_{1,0}(\ub{l}) - E_{0,1}(\ub{l} - \ub{k}))x^+} \over E_{1,0}(\ub{k}) -
E_{1,0}(\ub{l}) + E_{0,1}(\ub{l} - \ub{k})} \left( {-{\bbox
\epsilon}_{\perp,2s}\cdot \senk{k}\over k^+ \sqrt{l^+ - k^+}} + {-{\bbox
\epsilon}_{\perp,-2s}^* \cdot \senk{l}\over l^+ \sqrt{l^+ - k^+}} \right)
d_{-s,\ub{l}}^\dagger a_{\ub{l} - \ub{k}}
\nonumber \\ 
-&&{1 \over \sqrt{8\pi^3}} g\int_0^{k^+} d\ub{l}  {e^{i
\ha(E_{1,0}(\ub{l}) + E_{0,1}(\ub{k} - \ub{l}))x^+} \over E_{1,0}(\ub{k})
- E_{1,0}(\ub{l}) - E_{0,1}(\ub{k} - \ub{l})} \left( {-{\bbox
\epsilon}_{\perp,2s}\cdot \senk{k}\over k^+ \sqrt{k^+ - l^+}} + {-{\bbox
\epsilon}_{\perp,-2s}^* \cdot \senk{l}\over l^+ \sqrt{k^+ - l^+}} \right)
d_{-s,\ub{l}}^\dagger a_{\ub{k} - \ub{l}}^\dagger
\nonumber \\ 
-&&{1 \over \sqrt{8\pi^3}} g\int d\ub{l}  {e^{i \ha(-
E_{1,0}(\ub{l}) + E_{0,1}(\ub{l} + \ub{k}))x^+} \over E_{1,0}(\ub{k}) +
E_{1,0}(\ub{l}) - E_{0,1}(\ub{l} + \ub{k})} \left( {-{\bbox
\epsilon}_{\perp,2s}\cdot \senk{k}\over k^+ \sqrt{k^+ + l^+}} + {-{\bbox
\epsilon}_{\perp,-2s}^* \cdot \senk{l}\over l^+ \sqrt{k^+ + l^+}} \right)
b_{s,\ub{l}} a_{\ub{l} + \ub{k}}^\dagger
\nonumber \\ 
-&&{1 \over \sqrt{16\pi^3}} mg\int_{k^+}^\infty d\ub{l}
{e^{i \ha(E_{1,0}(\ub{l}) - E_{0,1}(\ub{l} - \ub{k}))x^+} \over
E_{1,0}(\ub{k}) - E_{1,0}(\ub{l}) + E_{0,1}(\ub{l} - \ub{k})} \left( {1
\over k^+ \sqrt{l^+ - k^+}} + {1 \over l^+ \sqrt{l^+ - k^+}} \right)
d_{s,\ub{l}}^\dagger a_{\ub{l} - \ub{k}}
\nonumber \\ 
-&&{1 \over \sqrt{16\pi^3}} mg\int_0^{k^+} d\ub{l}  {e^{i
\ha(E_{1,0}(\ub{l}) + E_{0,1}(\ub{k} - \ub{l}))x^+} \over E_{1,0}(\ub{k})
- E_{1,0}(\ub{l}) - E_{0,1}(\ub{k} - \ub{l})} \left( {1 \over k^+
\sqrt{k^+ - l^+}} + {1 \over l^+ \sqrt{k^+ - l^+}} \right)
d_{s,\ub{l}}^\dagger a_{\ub{k} - \ub{l}}^\dagger
\nonumber \\ 
-&&{1 \over \sqrt{16\pi^3}} mg\int d\ub{l}  {e^{i \ha(-
E_{1,0}(\ub{l}) + E_{0,1}(\ub{l} + \ub{k})) x^+} \over E_{1,0}(\ub{k}) +
E_{1,0}(\ub{l}) - E_{0,1}(\ub{l} + \ub{k})} \left( {1 \over k^+ \sqrt{k^+
+ l^+}} - {1 \over l^+ \sqrt{k^+ + l^+}} \right) b_{-s,\ub{l}} a_{\ub{l}
+ \ub{k}}^\dagger \Biggr] \Biggr\}\,.	
\nonumber				
\end{eqnarray} 
The fields $\psi_+$ and $\phi$ are free fields:
\begin{eqnarray}
\psi_+(x^+,\ub{x})&=&{1\over\sqrt{16\pi^3}} \sum_{s} \int d\ub{k} \chi_s
\; \Bigl[ b_{s,\ub{k}} e^{-i(\ha E_{1,0}(\ub{k}) x^+ +
\ub{k}\cdot\ub{x})}+ d^\dagger_{-s,\ub{k}}e^{+i(\ha E_{1,0}(\ub{k}) x^+
+\ub{k}\cdot\ub{x})}\Bigr]\,,	
\\		
\phi(x^+,\ub{x})&=&{1\over\sqrt{16\pi^3}}  \int d\ub{q}
{1\over\sqrt{q^+}}\Bigl[ a_{\ub{q}} e^{-i(\ha E_{0,1}(\ub{q}) x^+ +
\ub{q}\cdot\ub{x})}+ a^\dagger_{\ub{q}} e^{+i(\ha E_{0,1}(\ub{q}) x^+
+\ub{q}\cdot\ub{x})}\Bigr].			
\end{eqnarray} 
Thus we have obtained solutions for all the independent
degrees of freedom.  By use of equations (\ref{eompsi}) and
(\ref{eometa}) we can reconstruct the Fermi fields.  One can then
evaluate the Fermi fields on the surface $t = 0$ and work out the
relation between the equal-time Fermi modes and the light-cone Fermi
modes just as we did above for the Bose field.

\section{Several Pauli--Villars Fields of the Same Type}
\label{sec:SeveralPV}

In general the UV regularization of a renormalizable theory
requires the introduction of more than one PV field of
each type.  For example, the full Yukawa theory requires two PV
fermions, and QCD requires several PV fermions for each color and
flavor~\cite{pfp}.  Our method is easily extendable to such cases.
We will illustrate this for the theory studied in Ref.~\cite{bhm3},
Yukawa theory without fermion loops regulated with three PV Bose fields.
The theory includes one (physical) Fermi field and four Bose fields: the
physical field, which we will call
$\phi_1$; two negative metric PV fields, which we will call $\phi_2$ and
$\phi_4$; and a positive metric PV field, which we will call $\phi_3$.
From these we define the following four zero-norm fields:
\begin{eqnarray}
    \zeta &\equiv & N(\xi_1\phi_1 + \xi_2\phi_2 + \xi_3\phi_3 +
\xi_4\phi_4)\,,  \\
    \zeta_1 &\equiv& N(\xi_3\phi_1 + \xi_4\phi_2 - \xi_1\phi_3 -
\xi_2\phi_4)\,,  \\
    \zeta_2 &\equiv & N(\xi_2\phi_1 - \xi_1\phi_2 + \xi_4\phi_3 -
\xi_3\phi_4)\,,  \\
    \zeta_3 &\equiv & N (\xi_4\phi_1 - \xi_3\phi_2 - \xi_2\phi_3 +
\xi_1\phi_4)\,.
\end{eqnarray} 
The $\xi_i$ are relative coupling strengths for the
different fields and are chosen to satisfy constraints that accomplish
designated cancellations. In particular, we have $\xi_1=1$ to retain $g$
as the ordinary bare coupling and $\sum_i (-1)^{i+1}\xi_i^2=0$ to give
the $\zeta$ fields zero norm.  The factor $N$ is chosen such that the
$\zeta$ fields have the following commutation relations:
\begin{eqnarray}
\Bigl[\zeta(x^+,\ub{x}),\del^+\zeta_2(x^+,\ub{x}^\prime)\Bigr]
&=&i\kd3(\ub{x}-\ub{x}^\prime)\,,  
\\					
\Bigl[\zeta_1(x^+,\ub{x}),\del^+\zeta_3(x^+,\ub{x}^\prime)\Bigr]
&=&i\kd3(\ub{x}-\ub{x}^\prime)\,.					
\end{eqnarray} 
All other commutators are zero.  The value of $N$ is then given by
\begin{equation} 
N=1/\sqrt{2(\xi_1\xi_2+\xi_3\xi_4)}\,.
\end{equation}

We make the mode expansions
\begin{eqnarray}
\psi_+(\ub{x})&=&{1\over\sqrt{16\pi^3}} \sum_{s} \int d\ub{k} \; \chi_s \;
\Bigl[ b_{s,\ub{k}} e^{-i\ub{k}\cdot\ub{x}}+
d^\dagger_{-s,\ub{k}}e^{+i\ub{k}\cdot\ub{x}}\Bigr]\,,  \\					
\zeta(\ub{x})&=&{1\over\sqrt{16\pi^3}}  \int d\ub{q} {1\over\sqrt{q^+}}\;
\Bigl[ a_{\ub{q}} e^{-i\ub{q}\cdot\ub{x}}+ a^\dagger_{\ub{q}}
e^{+i\ub{q}\cdot\ub{x}}\Bigr]\,,  
\\					
\zeta_i(\ub{x})&=&{1\over\sqrt{16\pi^3}}  \int d\ub{q}
{1\over\sqrt{q^+}}\; \Bigl[ a_{i,\ub{q}} e^{-i\ub{q}\cdot\ub{x}}+
a^\dagger_{i,\ub{q}} e^{+i\ub{q}\cdot\ub{x}}\Bigr]\,,	\quad i=1,2,3\,.				
\end{eqnarray} 
If we take the masses of the fields, $\phi_i$, to be
$\mu_i$, we find the operator $P^-$ to be of the form
\begin{equation} 
P^-\equiv P^-_{(0)}+gP^-_{(1)}+g^2P^-_{(2)}\,,
\end{equation} 
where
\begin{eqnarray} 
P^-_{(0)}=&& \sum_{s} \int d\ub{k}
\Bigl[ {\senk{k}^{\ths2}+m^2\over k^+}\Bigr]
\Bigl(b^\dagger_{s,\ub{k}}b_{s,\ub{k}}+d^\dagger_{s,\ub{k}}d_{s,\ub{k}}\Bigr)
\nonumber \\ +&& \int d{\ub{q}}
\Bigl[ {\senk{q}^{\ths2}\over q^+}\Bigr] (a^\dagger_{\ub{q}}a_{2\ub{q}} +
a^\dagger_{2\ub{q}}a_{\ub{q}} + a^\dagger_{1,\ub{q}}a_{3\ub{q}} +
a^\dagger_{3\ub{q}}a_{1\ub{q}})
\nonumber \\ 
+ && N^2\int d{\ub{q}} \Bigl[  {\xi_2^2\mu^2_1 -
\xi_1^2\mu^2_2 + \xi_4^2\mu^2_3 - \xi_3^2\mu^2_4
    \over  q^+}a^\dagger_{\ub{q}}a_{\ub{q}}  +{\xi_4^2\mu^2_1 -
\xi_3^2\mu^2_2 + \xi_2^2\mu^2_3 - \xi_1^2\mu^2_4
    \over  q^+}a^\dagger_{2\ub{q}}a_{2\ub{q}}
\nonumber \\ 
+&&{\xi_1^2\mu^2_1 - \xi_2^2\mu^2_2 + \xi_3^2\mu^2_3 -
\xi_4^2\mu^2_4
    \over  q^+}a^\dagger_{3\ub{q}}a_{3\ub{q}}  +{\xi_3^2\mu^2_1 -
\xi_4^2\mu^2_2 + \xi_1^2\mu^2_3 - \xi_2^2\mu^2_4
    \over  q^+}a^\dagger_{4\ub{q}}a_{4\ub{q}}
\nonumber \\ 
+&&{\xi_1\xi_2(\mu^2_1 + \mu^2_2) + \xi_3\xi_4(\mu^2_3 +\mu^2_4)
\over  q^+}(a^\dagger_{\ub{q}}a_{3\ub{q}} +
a^\dagger_{3\ub{q}}a_{\ub{q}})
\nonumber \\ 
+&&{\xi_2\xi_4(\mu^2_1 - \mu^2_3) - \xi_1\xi_3(\mu^2_2 -\mu^2_4)
\over  q^+}(a^\dagger_{\ub{q}}a_{2\ub{q}} +
a^\dagger_{2\ub{q}}a_{\ub{q}})
\nonumber \\ 
+&&{\xi_2\xi_3(\mu^2_1 - \mu^2_4) + \xi_1\xi_4(\mu^2_2 -\mu^2_3)
\over  q^+}(a^\dagger_{\ub{q}}a_{4\ub{q}} +
a^\dagger_{4\ub{q}}a_{\ub{q}})
\nonumber \\ 
+&&{\xi_1\xi_4(\mu^2_1 - \mu^2_4) + \xi_2\xi_3(\mu^2_2 +\mu^2_3)
\over  q^+}(a^\dagger_{3\ub{q}}a_{2\ub{q}} +
a^\dagger_{2\ub{q}}a_{3\ub{q}})
\nonumber \\ 
+&&{\xi_1\xi_3(\mu^2_1 - \mu^2_3) - \xi_2\xi_4(\mu^2_2 -\mu^2_4)
\over  q^+}(a^\dagger_{3\ub{q}}a_{4\ub{q}} +
a^\dagger_{4\ub{q}}a_{3\ub{q}})
\nonumber \\ 
+&&{\xi_3\xi_4(\mu^2_1 + \mu^2_2) + \xi_1\xi_2(\mu^2_3 +\mu^2_4)
\over  q^+}(a^\dagger_{2\ub{q}}a_{4\ub{q}} +
a^\dagger_{4\ub{q}}a_{2\ub{q}})
 \Bigr]\,,
\end{eqnarray} 
and
\begin{eqnarray} 
P^-_{(1)}={1\over \sqrt{8\pi^3}} &&\sum_{s} \int d\ub{k}
d\ub{l} d\ub{q}
\ths\ths{ ({\bbox{\e}_{\perp,2s}}^{\ths *}\cdot\senk{l}) \over
l^+\sqrt{q^+} }
\nonumber \\
\times\Bigl\{ &&b^\dagger_{s,\ub{k}}b_{-s,\ub{l}}a_{\ub{q}}
	\ths\kd3(\ub{k}-\ub{l}-\ub{q})
+b^\dagger_{s,\ub{k}}d^\dagger_{s,\ub{l}}a_{\ub{q}}
	\ths\kd3(\ub{q}-\ub{k}-\ub{l})  \nonumber \\
+&&b^\dagger_{s,\ub{k}}b_{-s,\ub{l}}a^\dagger_{\ub{q}}
	\ths\kd3(\ub{k}+\ub{q}-\ub{l})
-d^\dagger_{s,\ub{l}}d_{-s,\ub{k}}a_{\ub{q}}
	\ths\kd3(\ub{k}+\ub{q}-\ub{l})\nonumber \\
+&&d^\dagger_{-s,\ub{k}}b_{-s,\ub{l}}a^\dagger_{\ub{q}}
	\ths\kd3(\ub{k}+\ub{l}-\ub{q})
-d^\dagger_{s,\ub{l}}d_{-s,\ub{k}}a^\dagger_{\ub{q}}
	\ths\kd3(\ub{k}-\ub{l}-\ub{q}) \Bigr\}+{\rm h.c.}
\nonumber  \\ 
+{m\over \sqrt{16\pi^3}} &&\sum_{s} \int d\ub{k} d\ub{l} d\ub{q}
             \ths\ths{1\over l^+\sqrt{q^+}}
\nonumber \\
\times\Bigl\{ &&b^\dagger_{s,\ub{k}}b_{s,\ub{l}}a_{\ub{q}}
	\ths\kd3(\ub{k}-\ub{l}-\ub{q})
+d^\dagger_{-s,\ub{l}}b^\dagger_{s,\ub{k}}a_{\ub{q}}
	\ths\kd3(\ub{k}+\ub{l}-\ub{q})
\nonumber \\ 
+&&b^\dagger_{s,\ub{k}}b_{s,\ub{l}}a^\dagger_{\ub{q}}
	\ths\kd3(\ub{k}+\ub{q}-\ub{l})
+d^\dagger_{s,\ub{l}}d_{s,\ub{k}}a_{\ub{q}}
	\ths\kd3(\ub{k}+\ub{q}-\ub{l})
\nonumber \\ 
+&&d_{s,\ub{k}}b_{-s,\ub{l}}a^\dagger_{\ub{q}}
	\ths\kd3(\ub{k}+\ub{l}-\ub{q})
+d^\dagger_{s,\ub{l}}d_{s,\ub{k}}a^\dagger_{\ub{q}}
	\ths\kd3(\ub{k}-\ub{l}-\ub{q}) \Bigr\}+{\rm h.c.}
\end{eqnarray} 
In this case we do have four point interactions.  They are given by
\begin{eqnarray} 
P^-_{(2)}={1\over 16\pi^3} &&\sum_{s} \int d\ub{k}
d\ub{l} d\ub{p} d\ub{q}
\ths\ths{1\over\sqrt{p^+q^+}}
\nonumber \\
\times\Bigl\{ &&b^\dagger_{s,\ub{k}} b_{s,\ub{l}} a_{\ub{p}} a_{\ub{q}}
\ths{1\over k^+-p^+}
\ths\kd3(\ub{k}-\ub{l}-\ub{p}-\ub{q}) +b^\dagger_{s,\ub{k}}
d^\dagger_{-s,\ub{l}} a_{\ub{p}} a_{\ub{q}}
\ths{1\over k^+-p^+}
\ths\kd3(\ub{k}+\ub{l}-\ub{p}-\ub{q})
\nonumber \\ 
+&&b^\dagger_{s,\ub{k}} b_{s,\ub{l}} a^\dagger_{\ub{p}}a_{\ub{q}}
\ths{1\over k^+-q^+}
\ths\kd3(\ub{k}+\ub{p}-\ub{l}-\ub{q}) +b^\dagger_{s,\ub{k}}
d^\dagger_{-s,\ub{l}} a^\dagger_{\ub{p}} a_{\ub{q}}
\ths{1\over k^+-q^+}
\ths\kd3(\ub{k} + \ub{l} +\ub{p}-\ub{q} )
\nonumber \\ 
+&&b^\dagger_{s,\ub{k}} b_{s,\ub{l}} a^\dagger_{\ub{p}}a_{\ub{q}}
\ths{1\over k^++p^+}
\ths\kd3(\ub{k}+\ub{p}-\ub{l}-\ub{q}) +b^\dagger_{s,\ub{k}}
d^\dagger_{-s,\ub{l}} a^\dagger_{\ub{p}} a_{\ub{q}}
\ths{1\over k^++p^+}
\ths\kd3(\ub{k} + \ub{l} + \ub{p} - \ub{q} )
\nonumber \\ 
+&&b^\dagger_{s,\ub{k}} b_{s,\ub{l}} a^\dagger_{\ub{p}}a^\dagger_{\ub{q}}
\ths{1\over k^++p^+}
\ths\kd3(\ub{k} + \ub{p} + \ub{q}- \ub{l} ) +d^\dagger_{s,\ub{k}}
d_{s,\ub{l}} a_{\ub{p}} a_{\ub{q}} \ths{1\over l^++p^+}
\ths\kd3(\ub{k}-\ub{l}-\ub{p}-\ub{q})
\nonumber \\ 
-&&d_{s,\ub{k}} b_{-s,\ub{l}} a^\dagger_{\ub{p}} 
          a_{\ub{q}}\ths{1\over k^++q^+}
\ths\kd3(\ub{k} +\ub{l} + \ub{q} - \ub{p} ) +d^\dagger_{s,\ub{k}}
d_{s,\ub{l}} a^\dagger_{\ub{p}} a_{\ub{q}}
\ths{1\over l^++q^+}
\ths\kd3(\ub{k} +\ub{p}- \ub{l} - \ub{q} )
\nonumber \\ 
-&&d_{s,\ub{k}} b_{-s,\ub{l}} a^\dagger_{\ub{p}} a_{\ub{q}}
\ths{1\over k^+-p^+}
\ths\kd3(\ub{k} + \ub{l} +\ub{q} - \ub{p} ) +d^\dagger_{s,\ub{k}}
d_{s,\ub{l}} a^\dagger_{\ub{p}} a_{\ub{q}}
\ths{1\over l^+-p^+}
\ths\kd3(\ub{k}+\ub{p}-\ub{l}-\ub{q})
\nonumber \\ 
-&&d_{s,\ub{k}} b_{-s,\ub{l}} a^\dagger_{\ub{p}}a^\dagger_{\ub{q}}
\ths{1\over k^+-p^+}
\ths\kd3(\ub{k}+\ub{l}-\ub{p}-\ub{q}) +d^\dagger_{s,\ub{k}} d_{s,\ub{l}}
a^\dagger_{\ub{p}} a^\dagger_{\ub{q}}
\ths{1\over l^+-p^+}
\ths\kd3(\ub{k} + \ub{p} + \ub{q} - \ub{l} )
\Bigr\}\,.
\end{eqnarray} 
Again, if we define an index given by minus the number of
$a^\dagger$ type quanta in the state, we find that, for $\mu_1 = \mu_2 =
\mu_3 = \mu_4$, all the terms in $P^-$ either leave the index unchanged
or lower it.  The system is again triangular and easy to solve.  In the
present case, unlike the situation we found above with PV Fermi fields,
the eigenstates project onto an infinite number of sectors.
Nevertheless, the coefficients can be obtained recursively.

We shall illustrate this with one example.  If we define a new vertex
amplitude by
\begin{equation}
  X(\ub{k},\ub{p},\ub{q}) \equiv {1 \over 16\pi^3}{1 \over \sqrt{p^+ q^+}}
\left( {1 \over k^+ - p^+} \right)\,,
\end{equation} 
we find that the projection of the eigenstate whose
highest state is $\beta_{+,\ub{k}}^\dagger \ket{0}$, onto the sectors
with not more than two Bose quanta, is given by
\begin{eqnarray}
    &&\beta_{+,\ub{k}}^\dagger \ket{0}
\nonumber \\ 
&&+ gm\int_{0}^{k^+} d\ub{l} {U(\ub{k},\ub{l}) \over
E_{1,0}(\ub{k}) - E_{1,1}(\ub{l},\ub{k} - \ub{l})} b_{+,\ub{l}}^\dagger
a_{\ub{k} - \ub{l}}^\dagger \ket{0} + g\int_{0}^{k^+} d\ub{l}
{V(\ub{k},\ub{l}) \over E_{1,0}(\ub{k}) - E_{1,1}(\ub{l},\ub{k} -
\ub{l})} b_{-,\ub{l}}^\dagger a_{\ub{k} - \ub{l}}^\dagger \ket{0}
\nonumber \\ 
&& + g^2\int_{0}^{k^+} d\ub{p} \int_{0}^{k^+ -p^+ } d\ub{q}
\nonumber \\ 
&&{m^2 U(\ub{k},\ub{k} - \ub{p}) U(\ub{k} - \ub{p},\ub{k} -
\ub{p} - \ub{q}) + V(\ub{k},\ub{k} - \ub{p}) V(\ub{k} - \ub{p},\ub{k} -
\ub{p} - \ub{q}) + X(\ub{k},\ub{p},\ub{q}) \over E_{1,0}(\ub{k}) -
E_{1,2}(\ub{k} - \ub{p} - \ub{q},\ub{p},\ub{q})} b_{+,\ub{k} - \ub{p} -
\ub{q}}^\dagger a_{\ub{p}}^\dagger a_{\ub{q}}^\dagger \ket{0}
\nonumber \\ 
&& + g^2\int_{0}^{k^+} d\ub{p} \int_{0}^{k^+ - p^+ } d\ub{q}
\nonumber \\ 
&&{m U(\ub{k},\ub{k} - \ub{p}) V(\ub{k} - \ub{p},\ub{k} -
\ub{p} - \ub{q}) +  m V(\ub{k},\ub{k} - \ub{p}) U(\ub{k} - \ub{p},\ub{k}
- \ub{p} - \ub{q}) \over E_{1,0}(\ub{k}) - E_{1,2}(\ub{k} - \ub{p} -
\ub{q},\ub{p},\ub{q})} b_{-,\ub{k} - \ub{p} - \ub{q}}^\dagger
a_{\ub{p}}^\dagger a_{\ub{q}}^\dagger \ket{0}\,.
\end{eqnarray}

\section{Conclusions}  \label{sec:Conclusions}

If quantum field theories are regulated with PV fields, and the masses of
the PV fields are set equal to the masses of the physical fields, the
resulting theories can be solved explicitly.
The eigenstates are nontrivial but can be constructed either in
closed form or in terms of recursion relations.  The spectrum of these
theories is identical to that of the corresponding noninteracting theory.
However, many eigenstates have zero norm, and the $S$ matrix is trivial;
these properties make the fully degenerate case difficult to interpret 
physically, as should be expected in a theory that violates unitarity 
so severely.

We have also shown that, in some cases, exact operator solutions can be
obtained.  These solutions can be used to study those general
properties of quantum field theories which depend on covariance, but not
on unitarity.  These properties allow us to also investigate the relation
between the light-cone representation and the equal-time representation.
Since these theories have a highly structured equal-time
vacuum, they may provide insight into the nontrivial equal-time vacuum
structure of theories such as QCD.

The existence of the solution for arbitrary coupling constant, but for 
zero values of the mass differences $\delta$ and $\Delta$ 
(or their equivalents in more
complicated theories), opens the possibility of doing perturbation
expansions, not in the coupling constant, but in the mass differences
between the Pauli--Villars and physical hadrons, much like the case of
broken supersymmetry.  We have shown that the wave functions of theories
can be obtained in exact form for degenerate physical and PV
masses.  Thus, if evolution equations in the mass differences can be
derived, we know how to initialize the solutions.  We plan to consider
this approach in future work.

\acknowledgments 
This work was supported by the Department of Energy
through contracts DE-AC03-76SF00515 (S.J.B.),  
DE-FG02-98ER41087 (J.R.H.), and DE-FG03-95ER40908 (G.M.).

\appendix

\section*{Light-Cone Conventions}

We define light-cone coordinates by
\begin{equation} 
x^\pm \equiv x^0 \pm x^3\,,					
\end{equation} 
with the transverse coordinates $\senk{x}\equiv (x^1,x^2)$
unchanged.  Covariant four-vectors are written as \eg\ $x^\mu =
(x^+,x^-,\senk{x}),$ with the spacetime metric
\begin{equation} 
g^{\mu\nu}=\left(\matrix{0&2&0&0\cr
		   2&0&0&0\cr
		   0&0&-1&0\cr
		   0&0&0&-1\cr}\right)\,.					
\end{equation} 
Explicitly, we have
\begin{equation} 
x\cdot y=g_{\mu\nu}x^\mu y^\nu =
         \ha(x^+y^-+x^-y^+)-\senk{x}\cdot\senk{y}\,.			
\end{equation} 
We also make use of an underscore notation: for
position-space variables we write
\begin{equation}
\ub{x} \equiv (x^-,\senk{x})\,,					
\end{equation} 
while for momentum-space variables
\begin{equation}
\ub{k} \equiv (k^+,\senk{k})\,.					
\end{equation} 
Then the dot product becomes
\begin{equation}
\ub{k} \cdot\ub{x} \equiv\ha k^+x^- - \senk{k}\cdot\senk{x}\,.
\end{equation} 
Spatial derivatives are defined by
\begin{equation}
\del_+ \equiv {\del\over\del x^+}\ths,\qquad
\del_- \equiv {\del\over\del x^-}\ths,\qquad
\del_{i} \equiv {\del\over\del x^i} .					
\end{equation}

The gamma matrices $\g^\pm\equiv\g^0\pm\g^3=(\g^\mp)^\dagger$ satisfy the
familiar relation
\begin{equation}
\{\g^\mu,\g^\nu\}=2g^{\mu\nu}\,,					
\end{equation} 
with $g^{\mu\nu}$ the light-cone metric.  It is simple to
verify that the (hermitian) matrices
\begin{equation}
\L_\pm\equiv\ha \g^0\g^\pm					
\end{equation} 
satisfy
\begin{equation}
\L_\pm^2=\L_\pm\ths,\qquad\L_\pm\L_\mp=0\ths,\qquad\L_++\L_-=1\,,
\end{equation} 
so that they serve as projectors on spinor space.  In the
Dirac representation of the $\g$-matrices, $\L_+$ is given by
\begin{equation}
\L_+=\ha\left(\matrix{1&0&1&0\cr 0&1&0&-1\cr
		      1&0&1&0\cr 0&-1&0&1\cr}\right) ,				
\end{equation} 
which has two eigenvectors, both with eigenvalue $+1$\@.
They are
\begin{equation}
\chi_{+\ha}={1\over\sqrt2}\left(\matrix{ 1\cr0\cr1\cr0\cr}\right)\,,
\qquad\qquad
\chi_{-\ha}={1\over\sqrt2}\left(\matrix{ 0\cr1\cr0\cr-1\cr}\right)\,.
\end{equation} 
These serve as a convenient spinor basis for the expansion
of the field $\psi_+\equiv \L_+\psi$ on the light cone.


\end{document}